\documentclass[aps,pre,amssymb,amsfonts,a4paper,twoside,twocolumn]{revtex4-1}
\usepackage{amssymb,amsmath,bm}
\usepackage{graphicx}
\usepackage{enumerate}
\usepackage[utf8]{inputenc}
\usepackage{cancel}
\usepackage{color}
\usepackage[colorlinks,linkcolor=blue,urlcolor=blue,citecolor=blue]{hyperref}
\usepackage{float}
\usepackage{soul}
\begin{document}

\title{Heat fluctuations in a harmonic chain of active particles}

\author{Deepak Gupta$^{1,2}$ and David A. Sivak$^{2}$}
\address{\noindent {$^{1}$Dipartimento di Fisica `G. Galilei', INFN, Universit\`a di Padova, Via Marzolo 8, 35131 Padova, Italy}}
\address{\noindent {$^{2}$Department of Physics, Simon Fraser University, Burnaby, British Columbia V5A 1S6, Canada}}

\date{\today}
\begin{abstract}
One of the major challenges in stochastic thermodynamics is to compute the distributions of stochastic observables for small-scale systems for which fluctuations play a significant role. Hitherto much theoretical and experimental research has focused on systems composed of passive Brownian particles. In this paper, we study the heat fluctuations in a system of interacting active particles. Specifically we consider a one-dimensional harmonic chain of $N$ active Ornstein-Uhlenbeck particles, with the chain ends connected to heat baths of different temperatures. We compute the moment-generating function for the heat flow in the steady state. We employ our general framework to explicitly compute the moment-generating function for two example single-particle systems. Further, we analytically obtain the scaled cumulants for the heat flow for the chain. Numerical Langevin simulations confirm the long-time analytical expressions for first and second cumulants for the heat flow for a two-particle chain.
\end{abstract}
\maketitle
\section{Introduction}
\label{intro}
Nonequilibrium systems are ubiquitous~\cite{seifert-1,van,c-j-1,ret-1}. Examples include molecular motors, engines, bio-molecules, colloidal particles, and chemical reactions. In stark contrast to equilibrium counterparts~\cite{eq-st}, a general framework to understand nonequilibrium systems is still missing. 
In the last couple of decades, researchers have found general relations governing systems arbitrarily far from equilibrium, such as the {\it fluctuation relations}~\cite{seifert-1}, notably including transient and steady-state fluctuation theorems~\cite{ft-1,ft-2,ft-3,ft-4}, the Jarzynski work-free energy relation~\cite{c-j-2}, the Crooks work-fluctuation theorem~\cite{crooks}, and the Hatano-Sasa relation~\cite{HSR}. Recently, the {\it thermodynamic uncertainty relation} was established~\cite{tur-1}, essentially bounding the precision of arbitrary currents by the average entropy production. This points to promising applications to infer dissipation by measuring {\color{black}arbitrary currents~\cite{Roldan-infer,tur-2,tur-3,tur-4,tur-5}.} 

Although these relations are independent of specific system details, fluctuations of observables (heat, work, entropy production, particle current, efficiency, etc.) remain dependent on the choice of the system. The probability distribution $P(\mathcal{A},\tau)$ of an observable $\mathcal{A}$ at time $\tau$ in a system of interest carries full information about the fluctuations of $\mathcal{A}$. In the long-time limit, $P(\mathcal{A},\tau)$ is expected to have a large-deviation form~\cite{ldf}, $P(\mathcal{A},\tau)\asymp e^{\tau \mathcal{I}(\mathcal{A}/\tau)}$, where the symbol $\asymp$ implies logarithmic equality and the large-deviation function is
\begin{align}
\mathcal{I}(a)\equiv \lim_{\tau\to\infty} \dfrac{1}{\tau}\ln P(\mathcal{A}=a\tau,\tau) \ , \label{ldf-intro}
\end{align}
for $\mathcal{A}$ scaling linearly with the observation time $\tau$~\cite{ldf}. Unfortunately, exact calculation of the large-deviation function is only known for a few systems (see some examples in Refs.~\cite{lde-1,lde-2,lde-3,lde-4,lde-5,lde-6,lde-7,lde-8}).

Another class of nonequilibrium systems, {\it active matter}~\cite{am-1,am-2,am-3,am-4,am-5,am-6,am-7,am-8,fsc-1,am-11}, has attracted significant attention in recent years. The individual components of active matter independently consume energy from an internal source (in addition to the surrounding environment) and perform directed motion~\cite{mod}, thereby breaking time-reversal symmetry. Systems exhibiting active behavior include fish schools~\cite{fsc-2}, flocking birds~\cite{fl-0,fl-1} and rods~\cite{fl-2}, light-activated colloids~\cite{csf}, bacteria~\cite{ecoli,bac}, synthetic micro-swimmers~\cite{syn-1,syn-2,syn-3}, and motile cells~\cite{mot-c}. Several interesting observations have been made in different settings, for example clustering~\cite{cl-1,cl-2}, absence of a well-defined mechanical pressure~\cite{press}, motility-induced phase separation~\cite{mips}, and jamming~\cite{jam}. Research has focused on numerous quantitative features of active systems, e.g., transport properties in exclusion processes~\cite{sfd-1,sfd-2,sfd-3,sfd-4}, position distributions with~\cite{res-1,res-2} and without resetting~\cite{pd-1,pd-2,pd-3}, survival probability~\cite{surp}, mean squared displacement and position correlation functions~\cite{ak}, spatio-temporal velocity correlation functions~\cite{sptm}, arcsine laws~\cite{arc}, and the perimeter of the convex hull~\cite{conh}. 

Three predominant types of modeling are used to describe the motion of an individual active particle: (1) an active Brownian particle (ABP)~\cite{abp-rtp}, (2) a run-and-tumble particle (RTP)~\cite{abp-rtp}, and (3) an active Ornstein-Uhlenbeck particle (AOUP)~\cite{aoup}.  Recently, inspired by two bacterial species ({\it Myxococcus xanthus} and {\it Pseudomonas putida}), Santra et al.\ introduced a new scheme, a {\it direction-reversing active Brownian particle} (DRABP), to model bacterial motion~\cite{ion}. These different models differ in how they model self-propulsion. In this paper we study the simplest model, an AOUP, which nevertheless introduces several rich behaviors such as motility-induced phase separation~\cite{mips}, glassy dynamics~\cite{glass}, accumulation at walls~\cite{walls}, and has recently been used to understand distance from equilibrium~\cite{ldsbo-1,ldsbo-2,ldsbo-3} and time-reversal symmetry breaking~\cite{aoup} of active-matter systems.

A central concern in nonequilibrium physics is heat conduction through a system of interest, connected to two heat baths at different temperatures. According to Fourier's law, the local current is proportional to the local temperature gradient. Much research has studied the microscopic details of this picture in, for example, harmonic chains~\cite{ht-1,Fogedby} and lattices~\cite{ht-2,ht-3,ht-4,one-two-har-anhar}, anharmonic chains~\cite{non-li, ht-6} and lattices~\cite{ht-5,one-two-har-anhar}, disordered harmonic chains~\cite{ht-7}, a harmonic chain with alternating masses \cite{Fogedby-0}, elastically colliding unequal-mass particles~\cite{ht-8}, a free Brownian particle \cite{Visco}, and Brownian oscillators~\cite{ht-9,Fogedby-2}. We are unaware of any study of heat conduction in a system of interacting active particles.

In this paper, we quantify the effect of activity on heat-transport properties (both average and fluctuations of heat flow) in a one-dimensional chain of $N$ AOUPs connected by harmonic springs. In the steady state, we compute the long-time limit of the moment-generating function for heat flow using the formalism developed in Ref. \cite{ht-1}. We use our framework to show explicit derivations for the moment-generating functions in the long-time limit for two different one-particle systems~\cite{ht-9,apal}. We write analytical expressions for the first two cumulants of the heat flow (higher cumulants can be computed similarly). For a two-AOUP chain, we also compare the long-time analytical results with numerical simulations performed using Langevin dynamics.

The paper is organized as follows. In Sec.~\ref{model}, we present the model and discuss the steady-state joint distribution. In Sec.~\ref{FP-sec}, we formally derive the distribution of heat flow in the long-time limit. In Sec.~\ref{sec:z-lm}, we compute the characteristic function for heat flow in the long-time limit in the steady state. We apply our formalism to two different examples in Sec.~\ref{examp}. Using the characteristic function, we analytically compute cumulants for heat flow in Sec.~\ref{cum} and compare the analytical results for a chain of two particles with numerical simulations. Finally, we conclude in Sec.~\ref{summ}.

\section{Setup}
\label{model}
\begin{figure}
  \begin{center}
    \includegraphics[scale =0.06]{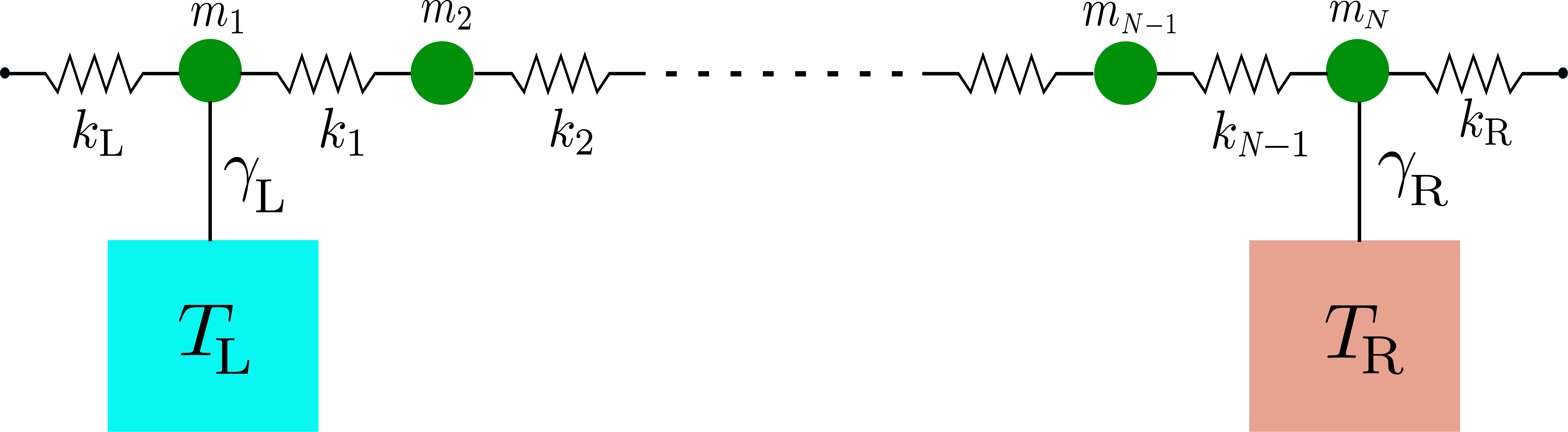}
    \caption{Schematic for one-dimensional harmonic chain of active Ornstein-Uhlenbeck particles coupled to two heat baths. $m_i$ is the mass of the $i$th particle, $k_i$ is the spring constant, and $T_{\rm L,R}$ and $\gamma_{\rm L,R}$, respectively, are temperatures of and friction coefficients coupling to the left (L) and right (R) heat baths.}
    \label{fig:scheme}
  \end{center}
\end{figure}

Consider a harmonic chain composed of $N$ active Ornstein-Uhlenbeck particles (AOUPs) in one dimension. Each particle is connected to its nearest neighbors with harmonic springs. Let $k_i$ be the stiffness constant of a spring connecting particles {\color{black}$i$ and $i+1$}. The left- and right-end particles, particles 1 and $N$, are connected to fixed locations with harmonic springs of stiffness $k_{\rm L}=k_0$ and  $k_{\rm R}=k_N$, respectively. Particles 1 and $N$ are, respectively, coupled with friction coefficients $\gamma_{\rm L}$ and $\gamma_{\rm R}$ to baths of temperature $T_{\rm L}$ and $T_{\rm R}$. Fig.~\ref{fig:scheme} shows a schematic of the system.

The underdamped dynamics of this coupled system obey, in matrix form, 
\begin{subequations}
\begin{align}
\dot X(t)&=V(t)\label{eq1}\\
M\dot V(t)&=-\Phi X-\Gamma V(t)+F(t)+B(t)\label{eq2}\\
\dot F(t)&=-R^{-1}F(t)+\mathcal{Z}(t)\label{eq3} \ ,
\end{align}
\end{subequations}
where the dot indicates a time derivative. In Eqs.~\eqref{eq1} and \eqref{eq2}, $X\equiv(x_1,x_2,\dots,x_N)^\top$, $V\equiv(v_1,v_2,\dots,v_N)^\top$, and $M\equiv\text{diag}(m_1,m_2,\dots,m_N)$, where $x_i$, $v_i$, and $m_i$, respectively, are the position, velocity, and mass of the $i$th particle. The left and right ends of the chain are connected to heat baths (see Fig.~\ref{fig:scheme}), so the noise vector is $B\equiv\delta_{i,1}\eta_{\rm L}+\delta_{i,N}\eta_{\rm R}=(\eta_{\rm L},0,\dots,0,\eta_{\rm R})^\top$, and the friction matrix is $\Gamma\equiv\delta_{i,j}(\delta_{i,1}\gamma_{\rm L}+\delta_{i,N}\gamma_{\rm R})$, where $\eta_{\rm L,{\rm R}}(t)$ are Gaussian thermal white noises with mean zero and correlations $\langle\eta_{\rm L}(t)\eta_{\rm L}(t')\rangle=2\gamma_{\rm L}T_{\rm L}\delta(t-t')$, $\langle\eta_{\rm R}(t)\eta_{\rm R}(t')\rangle=2\gamma_{\rm R}T_{\rm R}\delta(t-t')$, and $\langle\eta_{\rm L}(t)\eta_{\rm R}(t')\rangle=0$. For convenience, throughout the paper we set Boltzmann's constant to one. The nearest-neighbor coupling is reflected in the tridiagonal symmetric force matrix $\Phi$  with elements $\Phi_{i,j}\equiv (k_{i-1}+k_i)\delta_{i,j}-k_{i-1}\delta_{i,j+1}-k_i\delta_{i,j-1}$.
The chain particles are driven by force vector $F{\color{black}\equiv}(f_1,f_2,\dots, f_N)^\top$, with each {\it active force} $f_i$ dynamically evolving according to the Ornstein-Uhlenbeck (OU) equation in Eq.~\eqref{eq3}, with {\it active-noise} vector $\mathcal{Z}(t){\color{black}\equiv}(\zeta_1,\zeta_2,\dots,\zeta_N)^\top$, where each component $\zeta_i(t)$ is again a Gaussian white noise with mean zero and correlations $\langle\zeta_i(t)\zeta_j(t')\rangle=2D^{\rm a}_{i}\delta_{i,j}\delta(t-t')$.
The active and thermal noises are uncorrelated to each other, i.e., $\langle\eta_i(t)\zeta_j(t')\rangle=0$ for all $t,t'$. In Eq.~\eqref{eq3}, $R {\color{black}\equiv} {\rm diag}(t^{\rm a}_{1},t^{\rm a}_{2},\dots,t^{\rm a}_{N})$ is a diagonal matrix whose $(i,i)$-th element corresponds to the active relaxation time for the $i$th active force. The superscript `a' indicates {\it active}.

In the long-time stationary state, the mean of each active force $f_i$ is zero, with correlation 
\begin{align}
\langle f_i(t)f_i(t')\rangle=D^{\rm a}_{i}t^{\rm a}_{i}\exp\big\{-|t-t'|/t_{i}^{\rm a}\big\} \delta_{i,j}.
\end{align}
Notice that in the limit $t^{\rm a}_{i}\to 0$ and $D^{\rm a}_{i}\to\infty$ such that $(t^{\rm a}_{i})^2D^{\rm a}_{i}$ approaches a finite constant $2\mathcal{A}_{i}$, this active force is delta-correlated in time: $\langle f_i(t)f_j(t')\rangle=2\mathcal{A}_{i}~\delta_{i,j}\delta(t-t')$.

From Eqs.~\eqref{eq1}--\eqref{eq3}, the dynamical state vector $U{\color{black}\equiv}(X,V,F)^\top$ of the full system is linear with Gaussian white noises. Therefore, at {\color{black} a long time}, the distribution of $U$ reaches a stationary state (SS) Gaussian distribution (see Appendix~\ref{ss-app}){\color{black}:}
\begin{align}
P_{\rm SS}(U)\equiv \dfrac{1}{\sqrt{(2\pi)^{3N}\det[\Sigma]}}\exp\bigg[-\dfrac{1}{2}U^\top \Sigma^{-1}U\bigg],\label{ss-eqn}
\end{align} 
for correlation matrix
\begin{align}
\Sigma\equiv \langle U U^\top\rangle=\dfrac{1}{\pi}\int_{-\infty}^{+\infty}~d\omega~\bigg[\sum_{j=1}^{N}\dfrac{D^{\rm a}_{j} q_jq_j^\dagger}{\omega^2+(t^{\rm a}_{j})^{-2}}\\+\gamma_{\rm L}T_{\rm L}\ell_1\ell_1^\dagger+\gamma_{\rm R}T_{\rm R}\ell_N\ell_N^\dagger\bigg]\nonumber,
\end{align}
in which vectors $q_j$ and $\ell_j$, respectively, are 
\begin{subequations}
\begin{align}
q_j^\top&\equiv
(G_{1,j},G_{2,j},\dots,G_{N,j},i\omega_nG_{1,j},\label{eqs-7}\\&i\omega_nG_{2,j},\dots,i\omega_nG_{N,j},\delta_{1,j},\dots,\delta_{N,j})~~\text{for}~j=1,\dots, N,\notag\\
\ell_j^\top&\equiv(G_{1,j},G_{2,j},\dots,G_{N,j},i\omega_nG_{1,j},\label{eqs-8}\\&i\omega_nG_{2,j},\dots,i\omega_nG_{N,j}, \underbrace{0,0,\dots,0}_{N})~~\text{for}~ j=1,N.\notag
\end{align}
\end{subequations}
Notice that the symbol $\dagger$ refers to the combination of transpose and $\omega\to-\omega$ operations on a matrix. In both vectors~\eqref{eqs-7} and \eqref{eqs-8}, the first $N$ components, middle $N$ components, and final $N$ components, respectively, correspond to positions $x$, velocities $v$, and active forces $f$. $G_{i,j}(\omega)$ is the $(i,j)$-th matrix element of the symmetric Green's function matrix
\begin{align}
G(\omega)\equiv[\Phi-\omega^2 M+i\omega \Gamma]^{-1}.\label{gr-fn-def}
\end{align}

In this paper, we are interested in the fluctuations of heat flow from the left heat bath to the system~\cite{sekimoto} in a given time $\tau$ in the steady state $P_{\rm SS}(U_0)$ [see Eq.~\eqref{ss-eqn}]:
\begin{align}
Q_{\rm L}\equiv \int_0^\tau~dt~[\eta_{\rm L}(t)-\gamma_{\rm L} v_1(t)]v_1(t).\label{heat-eqn}
\end{align} 
In Sec.~\ref{sec:z-lm} we will show that the fluctuations of heat flow from the right heat bath can be computed using that of the left heat bath by applying suitable transformations.

Note that the above integral \eqref{heat-eqn} has to be interpreted with the Stratonovich rule~\cite{ito}. $Q_{\rm L}$ is not linear in the Gaussian state vector $U$, so we expect that its probability distribution $P(Q_{\rm L},\tau)$ is not generally Gaussian. 

In the following, we give a formal derivation of $P(Q_{\rm L},\tau)$ using the Fokker-Planck equation.

\section{Formal solution of the Fokker-Planck equation to derive $P(Q_{\rm L},\tau)$} 
\label{FP-sec}
To obtain the distribution of $Q_{\rm L}$, it is convenient to first compute the conditional characteristic function (also known as the conditional moment-generating function (CMGF)){\color{black}:}
\begin{align}
Z(\lambda,U,\tau|U_0)\equiv \int_{-\infty}^{+\infty}~dQ_{\rm L}~ e^{-\lambda Q_{\rm L}}~\rho(Q_{\rm L},U,\tau|U_0), \label{rhs-2}
\end{align}
where $\rho(Q_{\rm L},U,\tau|U_0)$ is the conditional joint distribution. We write the right-hand side as
\begin{align}
Z(\lambda,U,\tau|U_0)=
\bigg\langle e^{-\lambda Q_{\rm L}} \delta[U-U(\tau)]\bigg\rangle_{U_0},\label{eq-z}
\end{align}
where the angular brackets indicate averaging over all trajectories emanating from a fixed initial state vector $U_0$. Note that setting the conjugate variable $\lambda$ to zero in {\color{black}either} Eq.~\eqref{rhs-2} or \eqref{eq-z} gives the distribution $P(U,\tau|U_0)$ of state vector $U$ at time $\tau$ starting from a fixed initial vector $U_0$. $Z(\lambda,U,\tau|U_0)$ obeys the Fokker-Planck equation~\cite{van}{\color{black}:}
\begin{align}
\dfrac{\partial Z(\lambda,U,\tau|U_0)}{\partial \tau}=\mathcal{L}_\lambda Z(\lambda,U,\tau|U_0),\label{fp-eqn-1}
\end{align}
where $\mathcal{L}_\lambda$ is the Fokker-Planck operator {\color{black}[see Eq.~\eqref{FPO}]~\cite{van}.}

Since this differential equation is linear, the formal solution can be written as a linear combination of left- and right-eigenfunctions. In the long-time limit, the solution is dominated by the term corresponding to the largest eigenvalue $\mu(\lambda)$ of $\mathcal{L}_\lambda$, giving 
\begin{align}
Z(\lambda,U,\tau|U_0)\approx e^{\tau \mu(\lambda)} \chi(U_0,\lambda)\Psi(U,\lambda),\label{evec-form} 
\end{align}
where $\Psi(U,\lambda)$ is the corresponding right eigenfunction such that $\mathcal{L}_\lambda\Psi(U,\lambda)=\mu(\lambda)\Psi(U,\lambda)$, and $\chi(U_0,\lambda)$ is the projection of the initial state $U_0$ onto the left eigenvector corresponding to $\mu(\lambda)$.  Further note that the left- and right-eigenfunctions satisfy the normalization condition $\int dU~\chi(U,\lambda)\Psi(U,\lambda)=1$.  

Integrating the CMGF over both the steady-state distribution $P_{\rm SS}(U_0)$ of the initial state vector $U_0$~\eqref{ss-eqn} and the final state vector $U$ gives the characteristic function (moment-generating function){\color{black}:}
\begin{align}
Z(\lambda,\tau)\approx g(\lambda)e^{\tau \mu(\lambda)}, \label{f-z-l}
\end{align}
for prefactor
\begin{align}
g(\lambda)\equiv
\int dU_0~\int~dU~P_{\rm SS}(U_0) \chi(U_0,\lambda) \Psi(U,\lambda).\label{int-g-l}
\end{align}
Inverting $Z(\lambda,\tau)$ using the inverse Fourier transform gives the distribution function{\color{black}:}
\begin{align}
P(Q_{\rm L},\tau)&{\color{black}=} \dfrac{1}{2\pi i}\int_{-i\infty}^{+i\infty}~d\lambda~e^{\lambda Q_{\rm L}} Z(\lambda,\tau)\label{pdf-pq}\\&\approx\dfrac{1}{2\pi i}\int_{-i\infty}^{+i\infty}~d\lambda~g(\lambda)e^{\tau[ \mu(\lambda)+\lambda \mathcal{Q}]},\notag
\end{align}
where $\mathcal{Q} \equiv Q_{\rm L}/\tau$ is the  time-averaged heat rate entering the system from the chain's left end. The integral is performed along the vertical contour passing through the origin of the complex-$\lambda$ plane. 

When both $g(\lambda)$ and $\mu(\lambda)$ are analytic functions of $\lambda$, the {\color{black}integral \eqref{pdf-pq} can be approximated} (in the large-$\tau$ limit) using the saddle-point method~\cite{ldf}, giving the large-deviation form of the distribution
\begin{align}
P(Q_{\rm L}=\mathcal{Q} \tau,\tau)\asymp e^{\tau \mathcal{I}(\mathcal{Q})}, 
\end{align}
where $\mathcal{I}(\mathcal{Q})\equiv \mu(\lambda^*)+ \mathcal{Q}\lambda^*$ is the large-deviation function~\cite{ldf}, and $\lambda^*$ is the saddle point, a solution of
\begin{align}
\dfrac{\partial \mu(\lambda)}{\partial \lambda}\bigg|_{\lambda=\lambda^*(\mathcal{Q})} = -\mathcal{Q}. 
\end{align}
However, when $g(\lambda)$ has singularities in the region $\lambda \in [0,\lambda^*]$, special care is needed~\cite{ht-9,apal,apal-2}.

Computation of $\mu(\lambda)$ and $g(\lambda)$ using the Fokker-Planck equation is rather difficult. In Sec.~\ref{sec:z-lm}, we compute the characteristic function $Z(\lambda, \tau)$ using a method developed in \cite{ht-1} and previously used to compute distributions of quantities such as partial and apparent entropy productions~\cite{pep-1,pep-2,pep-3}, work fluctuations~\cite{apal,apal-2}, heat transport in lattices~\cite{ht-2}, and heat and work fluctuations for a Brownian oscillator~\cite{ht-9}. 

\begin{figure*}
    \centering
    \includegraphics[width=8.2cm]{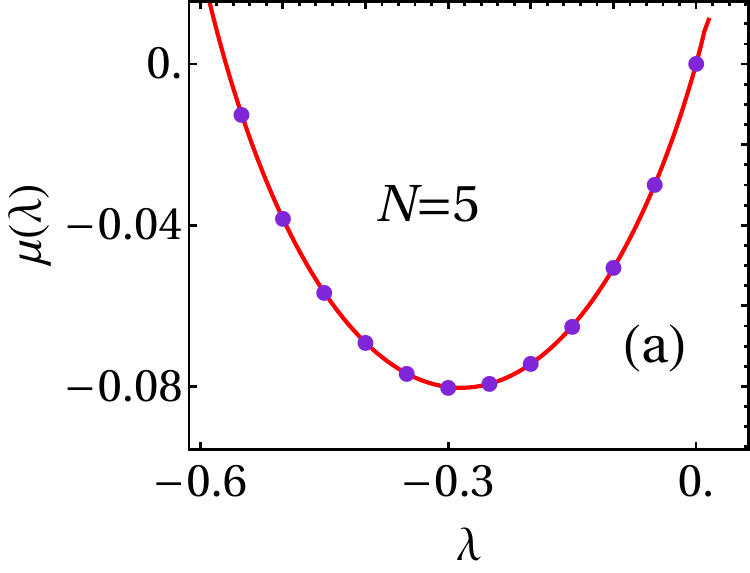}~~~~~~
    \includegraphics[width=8cm]{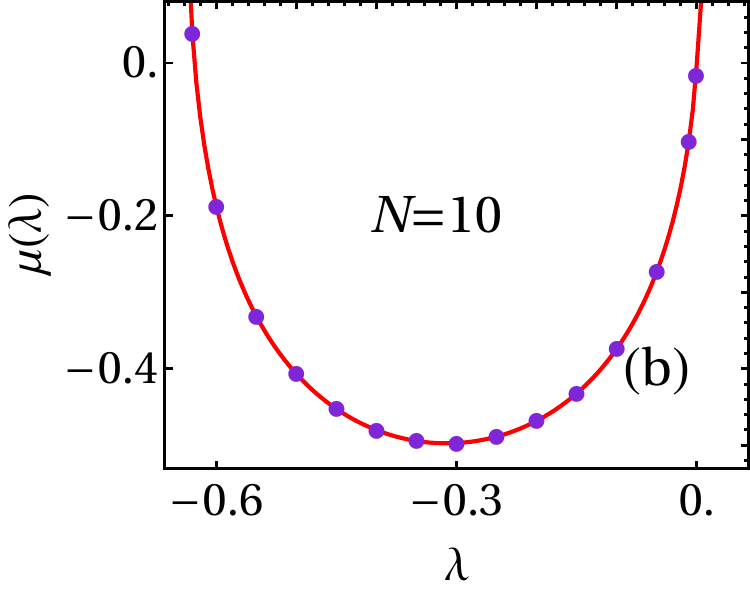}
    \caption{Largest eigenvalue $\mu(\lambda)$ as a function of $\lambda$ for $N=5$ (a) and $N=10$ (b). Circles: numerical computation of \eqref{mu-lam}. Solid red curve shows Eq.~\eqref{full-mu}. In both plots, $\gamma_{\rm L}=0.58,~\gamma_{\rm R}=0.75,~T_{\rm L}=1.58,~T_{\rm R}=0.11145$. For panel (a): $k_{\rm L}=0.1,~k_1=0.25,~k_2=0.2,~k_3=0.3,~k_4=0.4,~k_{\rm R}=0.5,~m_1=0.5,~m_2=0.27,~m_3=0.19,~m_4=1.48,~m_5=1.5,~D^{\rm a}_1=0.154,~D^{\rm a}_2=0.1254,~D^{\rm a}_3=0.147,~D^{\rm a}_4=0.3259,~D^{\rm a}_5=0.6548,~t^{\rm a}_{1}=1.154,~t^{\rm a}_{2}=2.2254,~t^{\rm a}_{3}=1.147,~t^{\rm a}_{4}=0.39,~t^{\rm a}_{5}=3.148.$ For panel (b): $m_\ell=0.1\ell,~t^{\rm a}_{\ell}=0.3\ell,~D^{\rm a}_{\ell}=0.5 \ell,~k_\ell=0.05(\ell+1), k_{\rm L}=0.05$, and~$k_{\rm R}=0.55$.}
    \label{fig:mu-lam}
\end{figure*}

\section{Computing the characteristic function $Z(\lambda,\tau)$}
\label{sec:z-lm}
In this section, we derive the largest eigenvalue $\mu(\lambda)$ and the prefactor $g(\lambda)$ appearing in the characteristic function $Z(\lambda,\tau)$ [see Eq.~\eqref{f-z-l}] for heat $Q_{\rm L}$ flowing through the left end of the $N$-particle system in the steady state. 

We first introduce the finite-time Fourier transform and its inverse~\cite{ht-1},
\begin{subequations}
\begin{align}
\tilde A(\omega_n)&\equiv \dfrac{1}{\tau}\int_0^\tau~dt~A(t)~e^{-i\omega_n t}\label{FT-1}\\
A(t)&\equiv \sum_{n=-\infty}^{+\infty}\tilde A(\omega_n)~e^{i\omega_n t}\label{FT-2},
\end{align}
\end{subequations}
where $\omega_n{\color{black}=} 2\pi n/\tau$ for integer $n$. 

We replace $\eta_{\rm L}(t)$ and $v_1(t)$ in the right-hand side (RHS) of Eq.~\eqref{heat-eqn} with their finite-time inverse Fourier transform representations~\eqref{FT-2}, and then integrate over time, obtaining the Fourier decomposition for the left heat flow:
\begin{align}
Q_{\rm L}= \dfrac{\tau}{2}\sum_{n=-\infty}^{+\infty} \big[\tilde \eta_{\rm L}(\omega_n)\tilde v_1(-\omega_n)+\tilde \eta_{\rm L}(-\omega_n)\tilde v_1(\omega_n)\label{FT-Q}\\-2\gamma_{\rm L} \tilde v_1(\omega_n)\tilde v_1(-\omega_n)\big]\notag. 
\end{align}
{\color{black}
We substitute the above expression of $Q_{\rm L}$ in the conditional characteristic function, $Z(\lambda,U,\tau|U_0)$, given in Eq.~\eqref{eq-z}, and compute the average (see Appendix~\ref{dd-eq} for detailed calculations), which eventually leads to (in the long-time limit)
\begin{subequations}
\begin{align}
Z(\lambda,U,\tau|U_0) &\approx \dfrac{e^{\tau \mu(\lambda)}e^{-\frac{1}{2} U^\top L_1 U}e^{-\frac{1}{2} U_0^\top L_2 U_0}}{\sqrt{(2\pi)^{3N}\det H_1(\lambda)}}  ,\label{z-u-u0}\\
\mu(\lambda)&\equiv-\dfrac{1}{4\pi} \int_{-\infty}^{+\infty}~d\omega~\ln [\det(\Lambda \Omega)],\label{mu-lam}\\
L_1(\lambda) &\equiv H_1^{-1}+H_1^{-1} H_2^\top, \\
L_2(\lambda) &\equiv -H_1^{-1}H_2^\top. \label{l1-l2}
\end{align}
\end{subequations}
Here $\mu(\lambda)$ is the largest eigenvalue of the Fokker-Planck operator $\mathcal{L}_\lambda$ \eqref{evec-form}, where in the integrand $\Lambda\equiv \frac{2}{\tau}~\text{diag}(D^{\rm a}_{1},D^{\rm a}_{2},\dots,D^{\rm a}_{N},\gamma_{\rm L} T_{\rm L},\gamma_{\rm R}T_{\rm R})$ is the noise correlation matrix appearing in the noise distributions in Eqs.~\eqref{n-d-1} and \eqref{n-d-0}, and {\color{black}$\Omega\equiv \Lambda^{-1}+\lambda \tau C$ (see Eq.~\eqref{cn-eqn} for $C$) \footnote{{\color{black}Here we converted the summation in the first term of Eq.~\eqref{gf} into an integral~\eqref{mu-lam} and thus dropped the subscript $n$ from the matrix $C_n$ given in \eqref{cn-eqn}.}}. The matrices} $H_1(\lambda)$, $H_2(\lambda)$, and $H_3(\lambda)$, respectively, are defined in Eqs.~\eqref{h1-eqn}, \eqref{h2-eqn}, and \eqref{h3-eqn}.}

Computation of the determinant in the integrand on the RHS of Eq.~\eqref{mu-lam} for arbitrary $N$ appears to us a difficult task. Nonetheless, for  $N= 1$ and $2$ (for $k\equiv k_1=k_{\rm L}=k_{\rm R}$) one can show that
\begin{widetext}
\begin{align}
\mu(\lambda)=-\dfrac{1}{4\pi}\int_{-\infty}^{+\infty}~d\omega~\ln\bigg[1+4\lambda(\Delta \beta-\lambda)\omega^2 \gamma_{\rm L} \gamma_{\rm R} T_{\rm L}T_{\rm R} |G_{1,N}|^2
-4\lambda(1+T_{\rm L}\lambda)\gamma_{\rm L}\sum_{\ell=1}^{N}\dfrac{D^{\rm a}_{\ell}|G_{1,\ell}|^2}{1+(\omega t^{\rm a}_{\ell})^{-2}}\bigg],\label{full-mu}
\end{align}
\end{widetext}
{\color{black}for $\Delta\beta \equiv T_{\rm R}^{-1}-T_{\rm L}^{-1}$.}
Further, Fig.~\ref{fig:mu-lam} shows the indistinguishability of \eqref{full-mu} and \eqref{mu-lam} for $N=5$ and $10$. Thus, we hypothesize that \eqref{full-mu} is valid for any $N$.

When the particles composing the chain have no activity ($D^{\rm a}_{\ell}\to 0$), we recover the same $\mu(\lambda)$ shown in Ref.~\cite{ht-1} for a harmonic chain of passive particles. Further, in the limit $t^{\rm a}_{\ell}\to 0$ and $D^{\rm a}_{\ell} \to \infty$ such that $(t^{\rm a}_{\ell})^2 D^{\rm a}_{\ell} \to 2\mathcal{A}_{\ell}$, the OU force $f_i$ is delta-correlated in time, and Eq.~\eqref{full-mu} becomes \begin{widetext} 
\begin{align}
\mu(\lambda)&=-\dfrac{1}{4\pi}\int_{-\infty}^{+\infty}~d\omega~\ln\bigg[1+4\lambda(\Delta \beta-\lambda)\omega^2 \gamma_{\rm L} \gamma_{\rm R} T_{\rm L}T_{\rm R} |G_{1,N}|^2
-8\lambda(1+T_{\rm L}\lambda)\gamma_{\rm L}  
\omega^2\sum_{\ell=1}^{N} \mathcal{A}_{\ell}|G_{1,\ell}|^2\bigg],\label{full-mu-2}
\end{align}
\end{widetext}
where the third term can be understood as the contribution coming from the Gaussian white noise with variance $2\mathcal{A}_{\ell}$ acting on the $\ell$th particle in the chain. In what follows, unless specified, we maintain the general case where $t^{\rm a}_{\ell}$ is positive and $D^{\rm a}_{\ell}$ is finite.

When the variable $\lambda$ conjugate to the heat $Q_{\rm L}$ is set to zero in $Z(\lambda,U,\tau|U_0)$ in Eq.~\eqref{z-u-u0}, both $\mu(\lambda)$ and $H_2(\lambda)$ vanish (see Eqs.~\eqref{mu-lam} and \eqref{h2-eqn}). This gives the distribution for $U$ at time $\tau$ starting from $U_0$, which in the large-$\tau$ limit approaches the (unique) steady state, 
\begin{align}
P_{\rm SS}(U)\equiv Z(0,U,\tau{\color{black}\to \infty}|U_0)=\dfrac{ e^{-\frac{1}{2} U^\top H^{-1}_1(0) U}}{\sqrt{(2\pi)^{3N}\det H_1(0)}}.\label{ss-eqn-2}
\end{align}
$H_1(0)$ can be obtained from Eq.~\eqref{h1-eqn} and shown equal to $M$, thereby recovering Eq.~\eqref{ss-eqn}.

Finally, we obtain the characteristic function $Z(\lambda,\tau)\equiv \langle e^{-\lambda Q_{\rm L}} \rangle$ by integrating $Z(\lambda,U,\tau|U_0)$ [see Eq.~\eqref{z-u-u0}] over the steady-state distribution $P_{\rm SS}(U_0)$ [see Eq.~\eqref{ss-eqn-2}] of the initial state vector $U_0$ and the final state vector $U$, thereby identifying the prefactor
\begin{align}
g(\lambda)= \big(\det[H_1(\lambda)L_1(\lambda)]\det[I+H_1(0)L_2(\lambda)]\big)^{-1/2},\label{g-lamb}
\end{align}
where $I$ is the identity matrix.

In this section, we calculated the characteristic function for the left heat flow $Q_{\rm L}(\tau)$ [Eq.~\eqref{heat-eqn}]. 
The characteristic function for the heat flow 
\begin{align}
Q_{\rm R}(\tau)\equiv \int_0^\tau~dt~[\eta_{\rm R}(t)-\gamma_{\rm R} v_N(t)]v_N(t),\label{heat-eqn-R}
\end{align}
from the right heat bath can be simply obtained from the characteristic function $Z(\lambda,\tau)$ for the left heat flow. This can be done by making the transformations: $\gamma_{\rm R}\longleftrightarrow \gamma_{\rm L},~T_{\rm R}\longleftrightarrow T_{\rm L}$ and relabelling the mass of particles, spring constants, strength of the active forces, and active-forces relaxation time, respectively, as
\begin{subequations}
\begin{align}
 &{(m_1,m_2,\dots,m_N)}\longrightarrow{(m_N,m_{N-1},\dots,m_1)},\\
 &{(k_{\rm L},k_1,\dots,k_{N-1},k_{\rm R})}\longrightarrow{(k_{\rm R},k_{N-1},\dots,k_{1},k_{\rm L})},\\
 &{(D_1^{\rm a},D_2^{\rm a},\dots,D_N^{\rm a})}\longrightarrow{(D_N^{\rm a},D_{N-1}^{\rm a},\dots,D_1^{\rm a})},\\
 &{(t_{1}^{\rm a},t_{2}^{\rm a},\dots,t_{N}^{\rm a})}\longrightarrow{(t_{N}^{\rm a},t_{N-1}^{\rm a},\dots,t_{1}^{\rm a})}.
\end{align}
\end{subequations}
In this way, $Q_{\rm L}$ exactly maps onto $Q_{\rm R}$. 
Applying these transformations to $Z(\lambda,\tau)$ gives the characteristic function $Z_{\rm R}(\lambda,\tau)\approx g_{\rm R}(\lambda)e^{\tau \mu_{\rm R}(\lambda)}$ for the heat flow from the right heat bath\footnote{The subscript R indicates the right heat bath.}, ultimately giving \begin{widetext}
\begin{align}
\mu_{\rm R}(\lambda)=-\dfrac{1}{4\pi}\int_{-\infty}^{+\infty}~d\omega~\ln\bigg[1-4\lambda(\Delta \beta+\lambda)\omega^2 \gamma_{\rm L} \gamma_{\rm R} T_{\rm L}T_{\rm R} |G_{1,N}|^2
-4\lambda(1+T_{\rm R}\lambda)\gamma_{\rm R} \sum_{\ell=1}^{N}\dfrac{D^{\rm a}_{\ell}|G_{N,\ell}|^2}{1+(\omega t^{\rm a}_{\ell})^{-2}}\bigg].\label{full-mu-R}
\end{align}
\end{widetext}

\section{Examples}
\label{examp}
So far, we have described how to compute the characteristic function $Z(\lambda,\tau)\approx g(\lambda)e^{\tau \mu(\lambda)}$ for our $N$-particle system. As discussed above, its inversion \eqref{pdf-pq} gives the full probability distribution $P(Q_{\rm L},\tau)$ of heat fluctuations. In this section, we consider two simple illustrative examples demonstrating our method to exactly compute both $\mu(\lambda)$ and $g(\lambda)$ at steady state.

\subsection{Heat fluctuation for a harmonic oscillator}
First, we specialize our general model to a single particle without OU driving force (i.e., a passive particle). This recovers the model previously used in Ref.~\cite{ht-9} to study the heat and work fluctuations for a Brownian oscillator. This particle evolves according to~\cite{ht-9}
\begin{subequations}
\begin{align}
\dot x(t) &= v(t),\\
m\dot v(t) &= kx(t)-(\gamma_{\rm L}+\gamma_{\rm R})v(t)+\eta_{\rm L}(t)+\eta_{\rm R}(t),
\end{align}
\end{subequations}
for particle position $x$, velocity $v$, mass $m$, and trap stiffness $k$.

In this case, we aim to compute the characteristic function $Z(\lambda,\tau)\sim g(\lambda)e^{\tau \mu(\lambda)}$ for heat flow $Q_{\rm L}$ into the system from the left heat bath~\eqref{heat-eqn}. Let us first compute $\mu(\lambda)$ (see the general expression~\eqref{mu-lam} for $N$ particles) in which $\Omega=\Lambda^{-1}+\lambda \tau C$. Here the noise correlation matrix governing the noise distributions~(\ref{n-d-1},~\ref{n-d-0}) is $\Lambda=\frac{2}{\tau}\,\text{diag}(\gamma_{\rm L} T_{\rm L},\gamma_{\rm R} T_{\rm R})$. Upon identifying $\Phi=k$, $M=m$, and $\Gamma=\gamma_{\rm L}+\gamma_{\rm R}$ (see Sec.~\ref{model}), the Green's function~\eqref{gr-fn-def} becomes
\begin{align}
G=[k-m\omega^2+i\omega(\gamma_{\rm L}+\gamma_{\rm R})]^{-1}.\label{g-fun}
\end{align}
In this case, the matrix $C$ appearing in the Fourier decomposition of the heat flow $Q_{\rm L}$ (see Eq.~\eqref{Q-eqn-FT}) can be deduced from \eqref{cn-eqn} for the special case of $N=1$: 
\begin{align}
C= \begin{pmatrix}
\overbrace{i\omega(G-G^*)-2\gamma_{\rm L} \omega^2|G|^2}^{2\gamma_{\rm R} \omega^2|G|^2}&&-i\omega G^*-2\gamma_{\rm L} \omega^2|G|^2\\
i\omega G-2\gamma_{\rm L} \omega^2|G|^2&&-2\gamma_{\rm L} \omega^2|G|^2
\end{pmatrix},\label{cs}
\end{align}
where $|G|^2\equiv G G^*$, and the first diagonal element of the matrix is re-written using a relation derived from Eq.~\eqref{g-fun},
\begin{align}
G-G^*=-2i\omega (\gamma_{\rm L}+\gamma_{\rm R}) |G|^2 \ .
\end{align}

Thus, $\mu(\lambda)$ given in Eq.~\eqref{mu-lam} becomes  
\begin{subequations}
\begin{align}
\mu(\lambda)&=-\dfrac{1}{4\pi}\int_{-\infty}^{+\infty}~d\omega~\ln [1+4\omega^2 |G|^2\gamma_{\rm L}\gamma_{\rm R}T_{\rm L}T_{\rm R}\lambda(\Delta\beta-\lambda)],\\
&=\dfrac{\gamma_{\rm L}+\gamma_{\rm R}}{2m}[1-\nu(\lambda)],\label{mu-sanjib} 
\end{align}
\end{subequations}
where {\color{black}again} $\Delta\beta\equiv T_{\rm R}^{-1}-T_{\rm L}^{-1}$, and
\begin{align}
\nu(\lambda)&\equiv\sqrt{1+4\frac{\gamma_{\rm L}\gamma_{\rm R}}{(\gamma_{\rm L}+\gamma_{\rm R})^2}T_{\rm L}T_{\rm R}\lambda(\Delta \beta-\lambda)}.\label{nu-lam}
\end{align}
To compute the prefactor $g(\lambda)$, we compute the three matrices $H_{1}(\lambda)$, $H_{2}(\lambda)$, and $H_{3}(\lambda)$ appearing in Eq.~\eqref{this-eqn-1} in exponents of the CMGF, $Z(\lambda,\tau,U|U_0)$. With the identification of
\begin{subequations}
\begin{align}
K_1&=(\ell_1^\top,\ell_2^\top)^\top,\\ 
K_2^\dagger&=(\ell_1^*,\ell_2^*),\\
\ell_1&=\ell_2, \\ 
a_1^\top&=[(1+2i\gamma_{\rm L} \omega G^*)\mathcal{R},2i\gamma_{\rm L}\omega G^*\mathcal{R}], \\ 
a_2&=[(1-2i\gamma_{\rm L}\omega G)\mathcal{R}^\dagger,-2i\gamma_{\rm L}\omega G\mathcal{R}^\dagger]^\top,\\
\mathcal{R}&= G(k,-im\omega)^\top,
\end{align}
\end{subequations}
$H_1(\lambda)$ given in Eq.~\eqref{h1-eqn} becomes
\begin{subequations}
\begin{align}
H_1(\lambda)&=\dfrac{\tau}{2\pi}\int_{-\infty}^{+\infty}~d\omega~\dfrac{\Omega_{22}-\Omega_{21}-\Omega_{12}+\Omega_{11}}{\det[\Omega]}\ell_1^*\ell_1^\top\label{h1-1-eqn}\\
&=\dfrac{\gamma_{\rm L}T_{\rm L}+\gamma_{\rm R}T_{\rm R}}{\pi}\int_{-\infty}^{+\infty}~d\omega~\times\label{h1-2-eqn}\\&\dfrac{|G|^2}{1+4\omega^2 |G|^2\gamma_{\rm L}\gamma_{\rm R}T_{\rm L}T_{\rm R}\lambda(\Delta\beta-\lambda)}\begin{pmatrix}
1&&i\omega\\
-i\omega&&\omega^2
\end{pmatrix}.\notag
\end{align}
\end{subequations}
In Eq.~\eqref{h1-1-eqn}, $\Omega_{ij}$ is the $(i,j)$-th matrix element of $\Omega$. The second line is obtained using the relations $\Omega_{22}-\Omega_{21}-\Omega_{12}+\Omega_{11}=\frac{\tau}{2}[(\gamma_{\rm L}T_{\rm L})^{-1}+(\gamma_{\rm R}T_{\rm R})^{-1}]$ and $\det \Omega=\frac{\tau^2}{4 \gamma_{\rm L}\gamma_{\rm R}T_{\rm L}T_{\rm R}}[1+4 \gamma_{\rm L}\gamma_{\rm R}T_{\rm L}T_{\rm R} \omega^2 |G|^2\lambda(\Delta \beta-\lambda)]$. The integrals of the off-diagonal elements vanish because the integrands are odd. Integrating the diagonal elements gives
\begin{align}
H_1(\lambda)=\dfrac{\gamma_{\rm L}T_{\rm L}+\gamma_{\rm R}T_{\rm R}}{(\gamma_{\rm L}+\gamma_{\rm R})\nu(\lambda)}
\begin{pmatrix}
k^{-1}&&0\\
0&&m^{-1}
\end{pmatrix}.
\end{align}

Similarly, $H_2(\lambda)$ (see Eq.~\eqref{h2-eqn}) becomes
\begin{subequations}
\begin{align}
H_2(\lambda)
&=\dfrac{\lambda}{\pi}\int_{-\infty}^{+\infty}d\omega~e^{-i\omega \epsilon}~\times\\&\dfrac{\gamma_{\rm L}T_{\rm L}(1+2i\gamma_{\rm L}\omega G^*)+2i\omega G^*\gamma_{\rm R}T_{\rm R}(\gamma_{\rm L}+\lambda \gamma_{\rm L}T_{\rm L})}{1+4\omega^2 |G|^2\gamma_{\rm L}\gamma_{\rm R}T_{\rm L}T_{\rm R}\lambda(\Delta\beta-\lambda)}\times\nonumber\\&\begin{pmatrix}k&&ik\omega\\-i\omega m&&m\omega^2\end{pmatrix}\nonumber\\
&=\frac{\lambda \gamma_{\rm L}T_{\rm L}-\frac{1}{2} (\gamma_{\rm L}+\gamma_{\rm R})  [\nu(\lambda)-1]}{(\gamma_{\rm L}+\gamma_{\rm R})\nu(\lambda )}\begin{pmatrix}1&&0\\0&&1\end{pmatrix},
\end{align}
\end{subequations}
and $H_3(\lambda)$ (see Eq.~\eqref{h1-eqn}) becomes
\begin{subequations}
\begin{align}
H_3(\lambda)
&=\dfrac{\lambda}{2\pi}\int_{-\infty}^{+\infty}d\omega~\dfrac{\mathcal{R}\mathcal{R}^\dagger}{\det(\Omega)}\times\\&\bigg[\lambda\tau^2\bigg(\dfrac{1}{2\gamma_{\rm R}T_{\rm R}}-2\gamma_{\rm L} \omega^2 |G|^2(\Delta \beta-\lambda)\bigg)+2\gamma_{\rm L} \det[\Omega]\bigg]\nonumber\\
&=\dfrac{\lambda(\gamma_{\rm L}+\lambda \gamma_{\rm L}T_{\rm L})}{\pi}\int_{-\infty}^{+\infty}d\omega~\times\\&\dfrac{|G|^2}{1+4\omega^2 |G|^2\gamma_{\rm L}\gamma_{\rm R}T_{\rm L}T_{\rm R}\lambda(\Delta\beta-\lambda)}\begin{pmatrix}k^2&&i\omega mk\\-i\omega mk&&m^2\omega^2\end{pmatrix}\nonumber\\
&=\dfrac{\lambda(\gamma_{\rm L}+\lambda \gamma_{\rm L}T_{\rm L})}{(\gamma_{\rm L}+\gamma_{\rm R}) \nu(\lambda)}\begin{pmatrix}k&&0\\0&&m\end{pmatrix}.
\end{align}
\end{subequations}
One can check that these matrices satisfy the condition $H_3(\lambda)=[I+H_2(\lambda)]H_1^{-1}(\lambda)H_2^\top(\lambda)$, ensuring the factorization of the CMGF into the product of factors that respectively capture the entire dependence on $U_0$ and $U$ (see Eqs.~\eqref{evec-form} and \eqref{z-u-u0}). Substituting $H_1(\lambda)$ and $H_2(\lambda)$ in $L_1(\lambda)$ and $L_2(\lambda)$ given in Eq.~\eqref{l1-l2} and Eq.~\eqref{g-lamb} gives
\begin{align}
g(\lambda)=\dfrac{4\nu(\lambda)}{[1+\nu(\lambda)]^2-[2\lambda \gamma_{\rm L}T_{\rm L}(\gamma_{\rm L}+\gamma_{\rm R})^{-1}]^2}.\label{g-lam-sanjib}
\end{align}

Finally, we write the characteristic function $Z(\lambda,\tau)\approx g(\lambda) e^{\tau \mu(\lambda)}$ (see Eq.~\eqref{f-z-l}) using Eqs.~\eqref{mu-sanjib} and \eqref{g-lam-sanjib}. Using the inverse transform \eqref{pdf-pq}, one can find the distribution of $P(Q_{\rm L},\tau)$ as discussed in Ref.~\cite{ht-9}.

\subsection{Work fluctuations for a Brownian particle driven by a correlated external random force} Here we specialize our general model to $N=1$, $T_{\rm L} =T_{\rm R} =T$, and $k_i=0$. The equations of motion for the particle read~\cite{apal}
\begin{subequations}
\begin{align}
\dot v(t)&=-\dfrac{1}{t_\gamma} v(t)+\dfrac{1}{m}f(t)+\dfrac{1}{m}\eta(t), \label{underdamped}\\
\dot f(t)&=-\dfrac{1}{t^{\rm a}}f(t)+\zeta(t),
\end{align} 
\end{subequations}
where $t_\gamma\equiv m/\gamma$ is the characteristic relaxation timescale of a particle's velocity, $\eta(t)$ is Gaussian thermal white noise of mean zero and correlation $\langle\eta(t)\eta(t') \rangle=2\gamma T\delta(t-t')$, and $f(t)$ is again an active OU force with mean zero and correlation $\langle f(t)f(t')\rangle=D^{\rm a}t^{\rm a}e^{-|t-t'|/t^{\rm a}}$ (see Sec.~\ref{model}). 
We use our extended framework to obtain the characteristic function for work done on the Brownian particle and show its consistency with previous calculations on this model by Pal and Sabhapandit~\cite{apal}.

The work due to external forcing is~\cite{sekimoto} (using the Stratonovich rule~\cite{ito})
\begin{align}
W\equiv \int_0^\tau~dt~f(t)v(t).
\end{align} 

Multiplying Eq.~\eqref{underdamped} by $v$ and integrating from 0 to $\tau$, the first law of thermodynamics reads
\begin{subequations}
\begin{align}
\dfrac{m}{2}(v_\tau^2-v_0^2)&=\int_0^\tau~dt~f(t) v(t)+\int_0^\tau~dt~[\eta(t)-\gamma v(t)] v(t)\\
\Delta \mathcal{U}&=W+Q,\label{f-law-2}
\end{align}
\end{subequations}
where the LHS, and the second term on the RHS, respectively, are the change in the internal energy and the heat flow from the bath to the system. Following Ref.~\cite{apal}, we define dimensionless work $\mathcal{W}\equiv \beta W$, where $\beta\equiv T^{-1}$ is the inverse temperature: the work is measured in units of thermal energy $k_{\rm B}T$ (Boltzmann's constant $k_{\rm B}$ is one).

With a suitable mapping, we compute the distribution of the dimensionless work $\mathcal{W}$. The CMGF (see Eq.~\eqref{eq-z}) for $\mathcal{W}$ can be written as
\begin{subequations}
\begin{align}
Z_{\mathcal{W}}(\lambda,U,\tau|U_0)&\equiv\bigg\langle e^{-\lambda \mathcal{W}} \delta[U-U(\tau)]\bigg\rangle_{U_0}\label{f-line}\\
&=e^{-\beta m\lambda (v_\tau^2-v_0^2)/2}\bigg\langle e^{\lambda \beta Q} \delta[U-U(\tau)]\bigg\rangle_{U_0}\label{s-line}\\
&=e^{-\beta m\lambda (v_\tau^2-v_0^2)/2}~Z(-\beta \lambda,U,\tau|U_0)\\ &\approx \dfrac{e^{\tau \mu(-\beta\lambda)}e^{-\frac{\beta m\lambda}{2} (v_\tau^2-v_0^2)}}{\sqrt{(2\pi)^{2}\det H_1(-\beta\lambda)}}\label{zw}\times\\&  ~~~~~~~~e^{-\frac{1}{2} U^\top L_1(-\beta\lambda) U}e^{-\frac{1}{2} U_0^\top L_2(-\beta\lambda) U_0}.\nonumber
\end{align}
\end{subequations}
The second line follows from the first law of thermodynamics~\eqref{f-law-2}. Further, we write the boundary contributions in the exponent in Eq.~\eqref{zw} in the matrix form:
\begin{align}
-\dfrac{\beta m\lambda}{2}(v_\tau^2-v_0^2){\color{black}=} \dfrac{1}{2}U^\top L_0(-\beta\lambda)U-\dfrac{1}{2}U_0^\top L_0(-\beta\lambda)U_0,\label{eq-zw-2} 
\end{align}
where $[L_0(-\beta \lambda)]_{i,j}\equiv-\beta m\lambda\delta_{i,1}\delta_{j,1}$, for $1\leq i,j\leq 2$. 

Substituting this in Eq.~\eqref{zw} yields
\begin{align}
Z_\mathcal{W}(\lambda,U,\tau|U_0)&\approx \dfrac{e^{\tau \mu(-\beta\lambda)}e^{-\frac{1}{2} U^\top L_3(-\beta\lambda) U}e^{-\frac{1}{2} U_0^\top L_4(-\beta\lambda) U_0}}{\sqrt{(2\pi)^{2}\det H_1\big(-\beta\lambda\big)}}  ,\label{zw-2}
\end{align}
where we have identified the modified exponents relating the exponents obtained from the CMGF of the (dimensionless) heat dissipated ($-\beta Q$) to the bath from the system, to that of the work on the particle by the external force: $L_3(-\beta\lambda)\equiv L_1(-\beta \lambda)-L_0(-\beta\lambda)$ and $L_4(-\beta\lambda)\equiv L_2(-\beta\lambda)+L_0(-\beta\lambda)$. We emphasize that the matrix $H_1(-\beta\lambda)$ corresponding to the work $\mathcal{W}$ remains the same as that of the heat dissipated to the bath $-\beta Q$.  

Integrating over the final state vector $U$ and the initial state vector $U_0$ with respect to the initial steady state distribution $P_{\rm SS}(U_0)$, gives the characteristic function (see Sec.~\ref{FP-sec}),
\begin{subequations}
\begin{align}
&Z_\mathcal{W}(\lambda,\tau)\approx g_\mathcal{W}(\lambda)e^{\tau\mu_\mathcal{W}(\lambda)},\label{zw-eqn-5} \\
&\mu_\mathcal{W}(\lambda)\equiv \mu(-\beta\lambda),\label{mu-w-eqn}\\
&g_\mathcal{W}(\lambda)\equiv \big(\det[H_1(-\beta\lambda)L_3(-\beta\lambda)]\big)^{-1/2}\times\label{gw-lam}\\&~~~~~~~~~~~~\big(\det[I+H_1(0)L_4(-\beta\lambda)]\big)^{-1/2}.\notag
\end{align}
\end{subequations}
Let us now compute $\mu_{\mathcal{W}}(\lambda)$ and $g_{\mathcal{W}}(\lambda)$. In this example, there is no harmonic confinement ($k=0$), so $\Phi=0$, $\Gamma=\gamma$, and $M=m$ (see Sec.~\ref{model}). Thus, the Green's function~\eqref{gr-fn-def} becomes $G=[i\omega \gamma-m\omega^2]^{-1}$. The diagonal matrix $\Lambda$ in the noise distributions in Eqs.~\eqref{n-d-1} and \eqref{n-d-0} is $\Lambda=\tfrac{2}{\tau}~\text{diag}(D^{\rm a},D)$. 
 
In the integrand of $\mu(\lambda)$ defined in Eq.~\eqref{mu-lam}, $\Omega= \Lambda^{-1}+\lambda\tau C$, where the Hermitian matrix $C$ can be obtained from Eq.~\eqref{cn-eqn} for one particle:
\begin{align}
C=\begin{pmatrix}
\dfrac{-2\gamma|G|^2}{1+(\omega^2t^{\rm a})^{-2}}&&\dfrac{G^*}{1+(i\omega t^{\rm a})^{-1}}\\
&&\\
\dfrac{-G}{-1+(i\omega t^{\rm a})^{-1}}&&0
\end{pmatrix}.
\end{align}
Substituting $\Lambda$ and $\Omega$ in $\mu(\lambda)$ in Eq.~\eqref{mu-lam} and making the transformation $\lambda\to -\beta \lambda$ gives
\begin{subequations}
\begin{align}
\mu_\mathcal{W}(\lambda)
&=-\dfrac{1}{4\pi}\int_{-\infty}^{+\infty}~d\omega~\ln\bigg[1+\dfrac{4\lambda(1-\lambda)\theta}{(\delta^2t_\gamma^2\omega^2+1)(t_\gamma^2\omega^2+1)}\bigg]\\
&=\dfrac{1}{2t_\gamma}[1-\bar\nu(\lambda)],\label{mu-w-fin}
\end{align}
\end{subequations}
for
\begin{subequations}
\begin{align}
\bar\nu(\lambda)&{\color{black}\equiv}\dfrac{1}{\delta}\left[\sqrt{1+\delta^2+2\delta\nu(\lambda)}-1\right]\\
\nu(\lambda)&{\color{black}\equiv}\sqrt{1+4\theta\lambda(1-\lambda)}.
\end{align}
\end{subequations}
Following~\cite{apal}, we introduced two dimensionless parameters, the relative strength $\theta\equiv (t^{\rm a})^2 D^{\rm a}/(\gamma T)$  of the external force with respect to thermal fluctuations, and the ratio $\delta\equiv t^{\rm a}/t_\gamma$ of the relaxation time of the external forcing and the relaxation time $t_\gamma$ of the particle's velocity. 

To compute $g_\mathcal{W}(\lambda)$, we first compute matrices $H_{1}(\lambda)$, $H_{2}(\lambda)$, and $H_{3}(\lambda)$ appearing in exponents of the CMGF $Z(\lambda,\tau,U|U_0)$ in Eq.~\eqref{this-eqn-1}, and then make the transformation $\lambda\to -\beta\lambda$ (see Eq.~\eqref{gw-lam}). In order to proceed further, we identify the following vectors which are helpful in computation of these matrices:
\begin{subequations}
\begin{align}
K_1 &= [(i\omega+1/t^{\rm a})^{-1}q_1^\top,\ell_1^\top]^\top{\color{black}\ ,}\\
K_2^\dagger &= [(-i\omega+1/t^{\rm a})^{-1}q_1^*,\ell_1^*]{\color{black}\ ,}\\
\ell_1^\top &= (i\omega G,0),\\
a_1^\top &= \left[\frac{2\gamma G^*{\mathcal{R}}}{-1+(i\omega t^{\rm a})^{-1}},(1+2i\gamma\omega G^*){\mathcal{R}}\right]{\color{black}\ ,}\\ 
\mathcal{R} &= [-i\omega mG,\dfrac{- G}{1+(i\omega t^{\rm a})^{-1}}]^\top,\\
a_2 &= \left[\frac{-2\gamma G\mathcal{R}^\dagger}{1+(i\omega t^{\rm a})^{-1}},(1-2i\gamma\omega G)\mathcal{R}^\dagger\right]{\color{black}\ ,}\\
q_1^\top &= (i\omega G,1)\ .
\end{align}
\end{subequations}

\begin{figure}
   \begin{center} 
   \includegraphics[width=8cm]{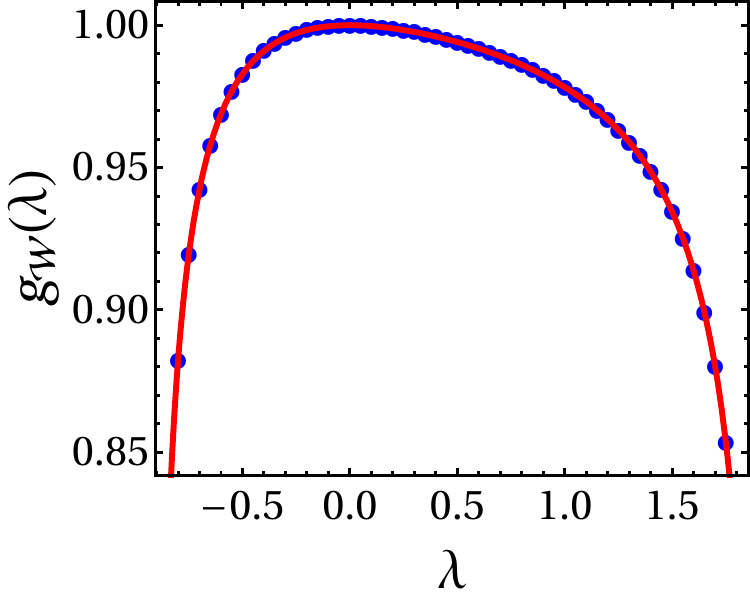}
   \caption{Numerical comparison of $g_{\mathcal{W}}(\lambda)$ (circles) given in Eq.~\eqref{gw-lam} and the prefactor (solid curve) in Eq.~(31) from Ref.~\cite{apal}. The parameters for the plot are $t_\gamma= 1.548,~\theta=0.1457,~\delta= 0.3484, ~m=0.14744,~T = 0.78$, and $\epsilon=10^{-4}$ in Eq.~\eqref{uh-2}. }
   \label{fig:g-comp}
   \end{center}
\end{figure}

Therefore, the matrices $H_1(\lambda)$, $H_2(\lambda)$, $H_3(\lambda)$, respectively given in Eqs.~\eqref{h1-eqn}, \eqref{h2-eqn}, and \eqref{h3-eqn}, can be simplified as\begin{widetext} 
\begin{subequations}
\begin{align}
H_1(\lambda)&=\dfrac{\tau}{2\pi}\int_{-\infty}^{+\infty}~\dfrac{d\omega}{\det[\Omega]}~\bigg[
\dfrac{\Omega_{22}q_1^*q_1^\top}{\omega^2+1/(t^{\rm a})^{2}}-\dfrac{\Omega_{21}\ell_1^*q_1^\top}{i\omega+1/t^{\rm a}}+\Omega_{11}\ell_1^*\ell_1^\top-\dfrac{\Omega_{12}q_1^*\ell_1^\top}{-i\omega+1/t^{\rm a}}\bigg],\label{uh-1}\\
H_2(\lambda)
&=\dfrac{\lambda\tau}{2\pi}\int_{-\infty}^{+\infty}\dfrac{d\omega~e^{-i\omega \epsilon}}{\det[\Omega]}~\bigg[\dfrac{2i\gamma\omega G^*\Omega_{22}\mathcal{R}q_1^\top}{\omega^2+1/(t^{\rm a})^{2}}-\dfrac{2\gamma G^*\Omega_{12}\mathcal{R}\ell_1^\top} {-1+(i\omega t^{\rm a})^{-1}}-\dfrac{(1+2i\omega\gamma G^*)\Omega_{21}\mathcal{R}q_1^\top}{i\omega+1/t^{\rm a}}
 +(1+2i\gamma\omega G^*)\Omega_{11}\mathcal{R}\ell_1^\top\bigg],\label{uh-2}\\
H_3(\lambda)
&=\dfrac{\lambda}{2\pi}\int_{-\infty}^{+\infty}\dfrac{d\omega}{\det[\Omega]}\bigg[\dfrac{4\gamma^2|G|^2}{1+(\omega^2t^{\rm a})^{-2}}\Omega_{22}-\dfrac{2\gamma G^*(1-2i\omega \gamma G)}{-1+(i\omega t^{\rm a})^{-1}}\Omega_{12} +\dfrac{2\gamma G(1+2i\omega \gamma G^*)}{1+(i\omega t^{\rm a})^{-1}}\Omega_{21}+\Omega_{11}+2\gamma\det[\Omega]\bigg]\mathcal{R}\mathcal{R}^\dagger.\label{uh-3}
\end{align}
\end{subequations}
\end{widetext}
Using the above integrals \eqref{uh-1}--\eqref{uh-3}, we verify the condition $H_3(\lambda)=[I+H_2(\lambda)]H_1^{-1}(\lambda)H_2^\top(\lambda)$ which ensures the factorization of the CMGF in terms of left and right eigenfunctions {\color{black}(see Appendix.~\ref{dd-eq})}. We substitute $H_1(\lambda)$ and $H_2(\lambda)$, respectively given in Eqs.~\eqref{uh-1} and \eqref{uh-2}, in $g_\mathcal{W}(\lambda)$ shown in \eqref{gw-lam}, and numerically compute the latter for a given parameters for comparison with the prefactor shown in Eq.~(31) of Ref.~\cite{apal}. Figure~\ref{fig:g-comp} shows that there is excellent agreement.
\begin{figure*}
  \begin{center}  
    \includegraphics[scale =0.68]{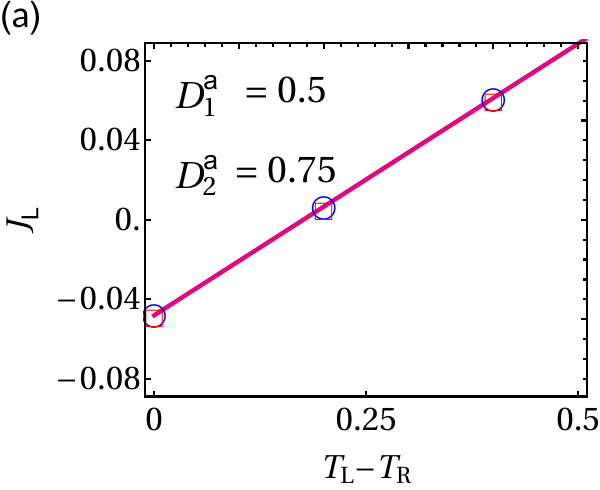}~~~~~~~~~~~
    \includegraphics[scale =0.68]{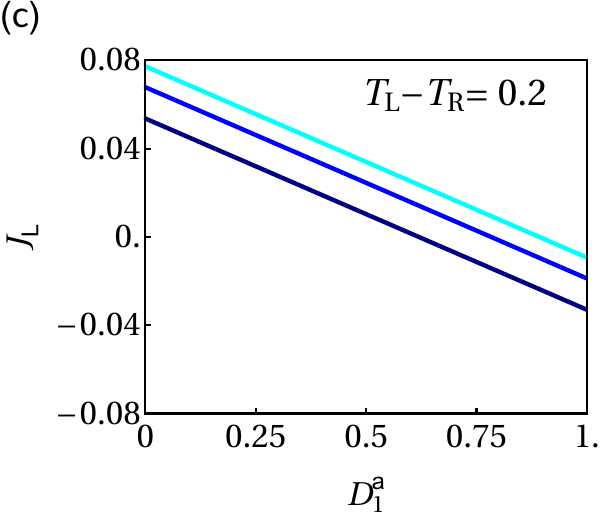}\\\bigskip
    \includegraphics[scale =0.68]{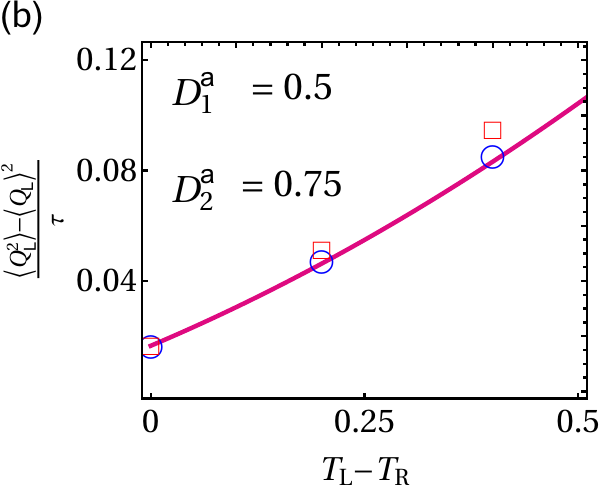}~~~~~~~~~
    \includegraphics[scale =0.68]{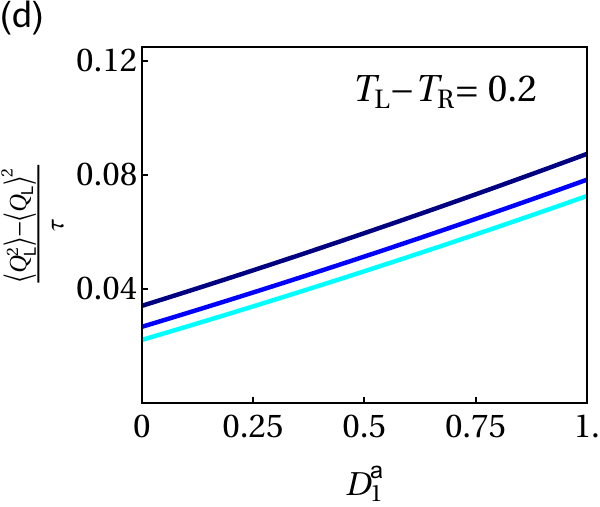}
    \caption{Cumulants of the left heat flow in the two-AOUP chain. (a) Analytical calculation (Eq.~\eqref{eq-cur}, solid curve) and numerical simulation {\color{black}(squares and circles, respectively, at $\tau=30$ and $\tau=300$)} of heat current for activities $D^{\rm a}_{1}=0.5$ and $D^{\rm a}_{2}=0.75$. (b) Analytical calculation (Eq.~\eqref{eq-var}, solid curve) and numerical simulation {\color{black}(squares and circles, respectively, at $\tau=30$ and $\tau=300$)} of scaled heat variance for $D^{\rm a}_{1}=0.5$ and $D^{\rm a}_{2}=0.75$. (c,d) Analytical results given in Eqs.~\eqref{eq-cur} and \eqref{eq-var}, respectively, as functions of activity $D^{\rm a}_{1}$, for $T_{\rm L}-T_{\rm R}=0.2$ and $D^{\rm a}_{2}=0.75$, $2.25$, and $4.5$ ({\color{black}curve color intensity} increases with $D_2^{\rm a}$). In all plots, $\gamma_{\rm L}=0.7$, $\gamma_{\rm R}=0.8$, $T_{\rm R}=0.1$, $t_{1}^{\rm a}=t_{2}^{\rm a}=0.15$, $m_1=0.1,~m_2=0.15$, $k_{\rm L}=0.75,~k_{1}=0.85,~k_{\rm R}=0.7$. Numerical simulations are performed for 
    $dt=10^{-4}$, and averaged over $10^4$ realizations.} 
    \label{cgf-fig}
  \end{center}
\end{figure*}
\begin{figure*}
  \begin{center}  
    \includegraphics[scale =0.68]{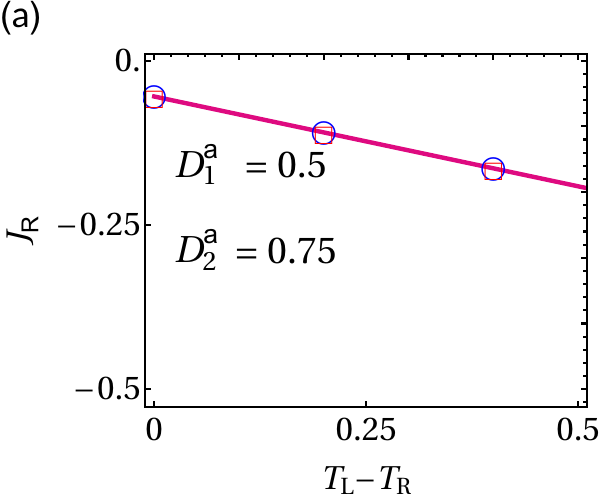}~~~~~~~~~~~
    \includegraphics[scale =0.68]{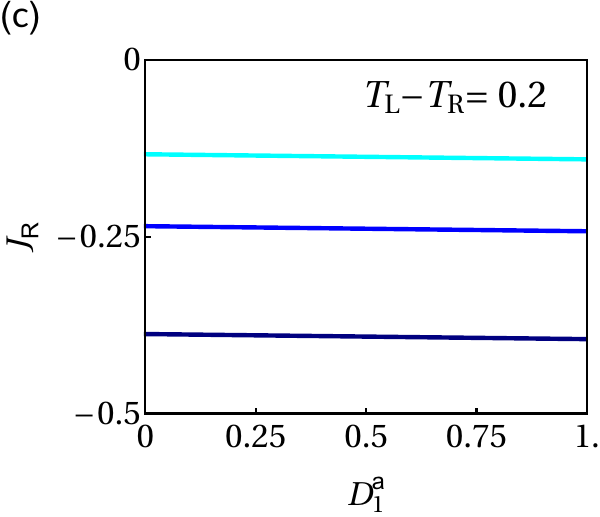}\\\bigskip
    \includegraphics[scale =0.68]{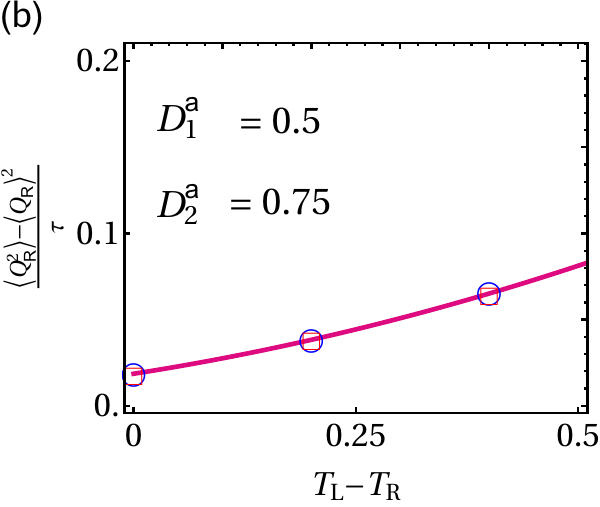}~~~~~~~~~
    \includegraphics[scale =0.68]{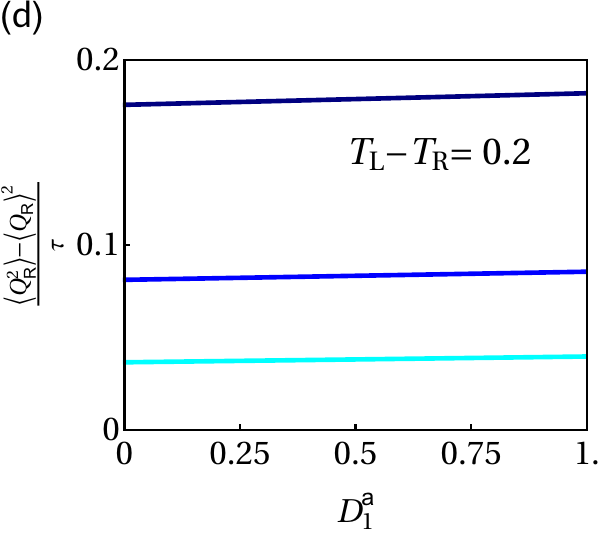}
    \caption{{\color{black}Cumulants of the right heat flow in the two-AOUP chain. (a) Analytical calculation (Eq.~\eqref{eq-cur-R}, solid curve) and numerical simulation (squares and circles, respectively, at $\tau=30$ and $\tau=300$) of heat current for activities $D^{\rm a}_{1}=0.5$ and $D^{\rm a}_{2}=0.75$. (b) Analytical calculation (Eq.~\eqref{eq-var-R}, solid curve) and numerical simulation (squares and circles, respectively, at $\tau=30$ and $\tau=300$) of scaled heat variance for $D^{\rm a}_{1}=0.5$ and $D^{\rm a}_{2}=0.75$. (c,d) Analytical results given in Eqs.~\eqref{eq-cur-R} and \eqref{eq-var-R}, respectively, as functions of activity $D^{\rm a}_{1}$, for $T_{\rm L}-T_{\rm R}=0.2$ and $D^{\rm a}_{2}=0.75$, $2.25$, and $4.5$ ({\color{black}curve color intensity} increases with $D_2^{\rm a}$). Other parameters are same as in Fig.~\ref{cgf-fig}.}}
    \label{cgf-fig-R}
  \end{center}
\end{figure*}
\begin{figure*}
    \centering
    \includegraphics[scale =.68]{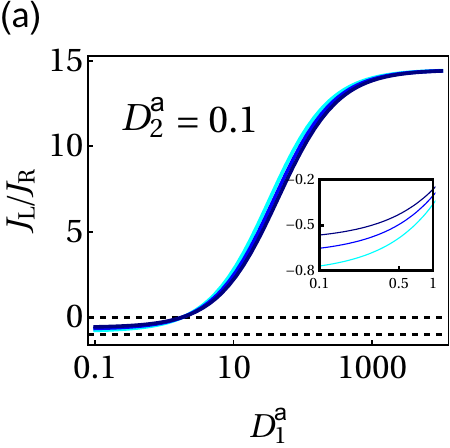}~~~~~
    \includegraphics[scale =0.68]{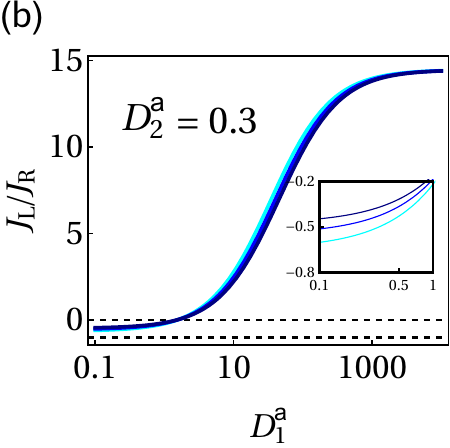}~~~~~
    \includegraphics[scale =0.68]{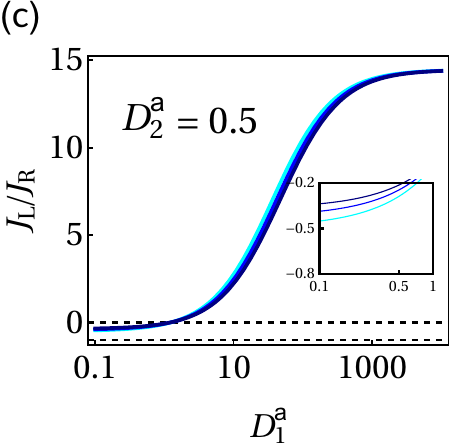}\\\bigskip
    ~~~~\includegraphics[scale=0.68]{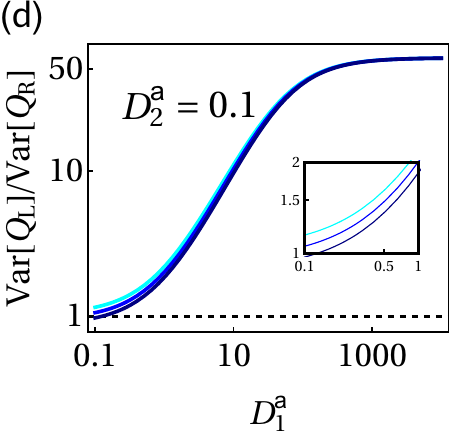}~~~~~~~~~~~~
    \includegraphics[scale =0.68]{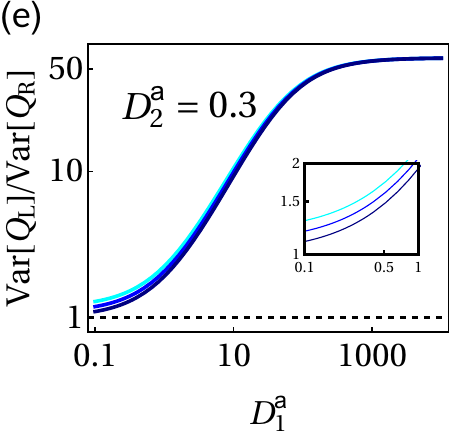}~~~~~~~~~~~~
    \includegraphics[scale =0.68]{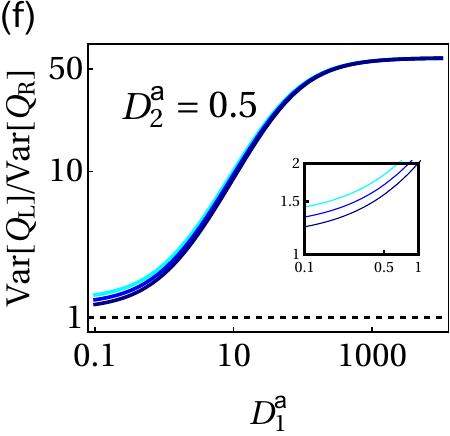}
    \caption{{\color{black}Ratio of left and right scaled 
    cumulants {\color{black}of heat flow} 
    for three-AOUP chain. (a-c) Ratio of the left~\eqref{eq-cur} and right heat currents~\eqref{eq-cur-R}, as a function of $D_1^{\rm a}$. (d-f) Ratio of variances of left heat flow~\eqref{eq-var} and of right heat flow~\eqref{eq-var-R}, as a function of $D_1^{a}$. 
    $D_3^{\rm a} = (0.1,~0.3,~0.5)$ ({\color{black}curve color intensity} increases with $D_3^{\rm a}$). {\color{black}Inset shows the zoom of the corresponding region of the main plot. Horizontal dashed lines in (a-c) correspond to $J_{\rm L}/J_{\rm R}=-1$ and $J_{\rm L}/J_{\rm R}=0$ and in (d-f) for ${\rm Var}[Q_{\rm L}]/{\rm Var}[Q_{\rm R}]=1$. }
    Throughout, $T_{\rm L}=1$, $T_{\rm R}=0.1$, $\gamma_{\rm L}=0.7$, $\gamma_{\rm R}=0.8$, $t^{\rm a}_{i}=0.15~\forall~i$, $m_1=0.5$, $m_2=0.3$, $m_3=0.2$, $k_{\rm L}=0.5$, $k_1=0.3$, $k_2=0.2$, and $k_{\rm R}=0.4$.} } 
    \label{fig:ratio}
\end{figure*}
\begin{figure}  
    \centering
    \includegraphics[scale =0.68]{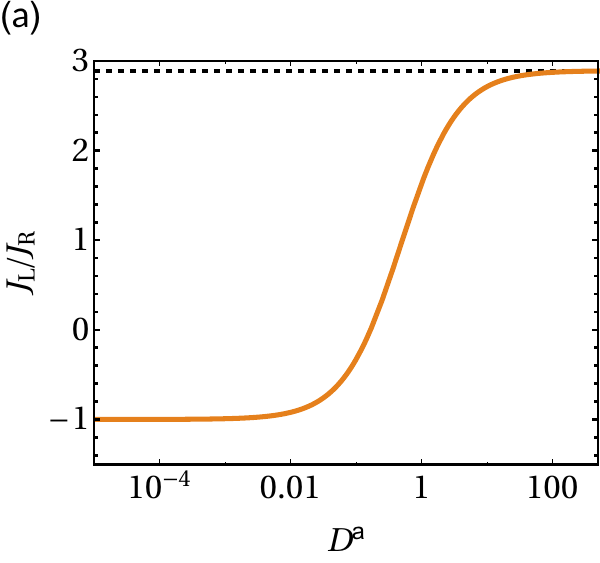}\\\bigskip
    \includegraphics[scale =0.78]{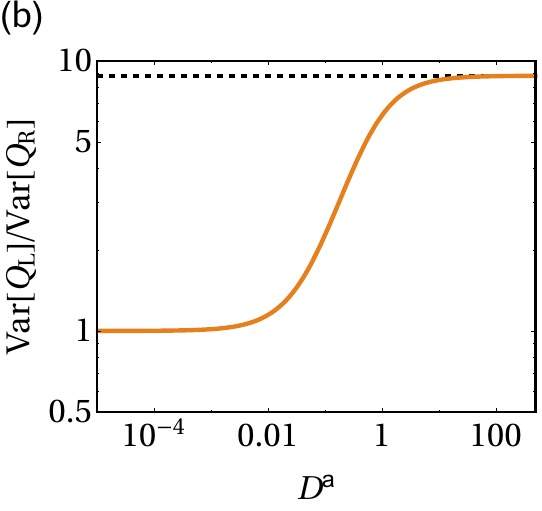}
    \caption{{\color{black}Ratios of left and right heat flow currents [Eqs.~\eqref{eq-cur} and \eqref{eq-cur-R}] and variances [Eqs.~\eqref{eq-var} and \eqref{eq-var-R}] for 4-AOUP chain with identical particle activities ($D_i^{\rm a} = D^{\rm a}~\forall~i$). Horizontal dashed lines: asymptotic values {\color{black}obtained from the dominating contributions of ratio of Eqs.~\eqref{eq-cur} and \eqref{eq-cur-R} for (a) and Eqs.~\eqref{eq-var} and \eqref{eq-var-R} for (b)} at large $D^{\rm a}$. 
    Throughout, $T_{\rm L}=1$, $T_{\rm R}=0.1$, $\gamma_{\rm L}=1.7$, $\gamma_{\rm R}=0.8$, $t^{\rm a}_{i}=0.15~\forall~i$, $m_1=0.2$, $m_2=0.2$, $m_3=0.3$, $m_4=0.5$, $k_{\rm L}=0.2$, $k_1=0.4$, $k_2=0.2$, $k_3=0.1$, and $k_{\rm R}=0.1$.}}
    \label{fig:ratio-same-d}
\end{figure}

Thus, we write the characteristic function $Z_\mathcal{W}(\lambda,\tau)$ (see Eq.~\eqref{zw-eqn-5}) using $\mu_\mathcal{W}(\lambda)$ and $g_\mathcal{W}(\lambda)$ given in Eqs.~\eqref{mu-w-fin} and \eqref{gw-lam}, respectively. One can invert the former using the inverse {\color{black}Fourier} transform defined in Eq.~\eqref{pdf-pq} and obtain $P(\mathcal{W},\tau)$ as discussed in Ref.~\cite{apal}.

Therefore, in this section, using two different examples, we have shown how our general framework can be employed to exactly calculate CMGFs for non-Gaussian observables in the long-time limit.

\section{Cumulants of heat flow}
\label{cum}
In Sec.~\ref{sec:z-lm}, we computed the characteristic function $Z(\lambda,\tau)$ for the heat entering the left end of the harmonic chain of $N$ AOUPs in the steady state. For a given number $N$ of particles, one can, in principle, compute both $\mu(\lambda)$ and $g(\lambda)$ as discussed in Sec.~\ref{examp}, and invert $Z(\lambda,\tau)\sim g(\lambda)e^{\tau \mu(\lambda)}$ using the inverse {\color{black}Fourier} transform~\eqref{pdf-pq} to give the full distribution for $Q_{\rm L}$. Since $Z(\lambda,\tau)$ is the moment-generating function, its logarithm gives the cumulant-generating function. In the long-time limit, 
\begin{align}
\dfrac{1}{\tau}\ln Z(\lambda,\tau)
=\dfrac{1}{\tau}\ln \langle e^{-\lambda Q_{\rm L}}\rangle=\mu(\lambda)+\dfrac{1}{\tau}\ln g(\lambda).\label{scgf}
\end{align}
If $g(\lambda)$ is an analytic function of $\lambda$, differentiating on both sides and setting $\lambda$ to zero gives the first scaled cumulant (mean) of the heat flow (i.e., the left heat current)
\begin{align}
-\dfrac{\partial \mu(\lambda)}{\partial \lambda}\bigg|_{\lambda=0} = \dfrac{\langle Q_{\rm L} \rangle}{\tau} \equiv J_{\rm L}.\label{cur-1}
\end{align} 
Similarly, differentiating twice and setting $\lambda=0$ gives
\begin{align} 
\dfrac{\partial^2 \mu(\lambda)}{\partial\lambda^2}\bigg|_{\lambda=0}=\dfrac{1}{\tau}(\langle Q_{\rm L} \rangle^2-\langle Q_{\rm L}^2 \rangle).\label{cur-2}
\end{align}
Higher-order cumulants can be obtained similarly. Notice that in the long-time limit, the contributions from $g(\lambda)$ are lower order in $\tau$ and so vanish in both Eqs.~\eqref{cur-1} and \eqref{cur-2}. This is even true if $g(\lambda)$ has singularities. For example, consider a case in which $g(\lambda)=\dfrac{g_0(\lambda)}{\prod_{i}(\lambda_i-\lambda)^{\alpha_i}}$, where $g_0(\lambda)$ is an analytic function of $\lambda$ and the singularities are on the right-side of the origin: $\lambda_i>0$. (Here $\alpha_i$ need not be integers.) Substituting in Eq.~\eqref{scgf}, in the long-time limit there is no contribution from $g(\lambda)$ in Eqs.~\eqref{cur-1} and \eqref{cur-2} (and similarly for higher cumulants).

Therefore, substituting $\mu(\lambda)$ (given in Eq.~\eqref{full-mu}) in Eqs.~\eqref{cur-1} and \eqref{cur-2}, the first two cumulants can be obtained in the integral form: \begin{widetext}
\begin{align}
J_{\rm L}&=\dfrac{T_{\rm L}-T_{\rm R}}{4\pi}\int_{-\infty}^{+\infty}~d\omega~4\gamma_{\rm L}\gamma_{\rm R} \omega^2 |G_{1,N}|^2-\dfrac{1}{4\pi}\int_{-\infty}^{+\infty}~d\omega~ 4\gamma_{\rm L}\sum_{\ell=1}^{N}\dfrac{D^{\rm a}_{\ell}|G_{1,\ell}|^2}{1+(\omega t^{\rm a}_{\ell})^{-2}},\label{eq-cur}\\
\dfrac{1}{\tau}[\langle Q_{\rm L}^2 \rangle-\langle Q_{\rm L} \rangle^2]&=\dfrac{1}{4\pi} \int_{-\infty}^{+\infty}~d\omega~\bigg[\bigg\{4\gamma_{\rm L}\gamma_{\rm R}(T_{\rm L}-T_{\rm R})\omega^2|G_{1,N}|^2-4\gamma_{\rm L}\sum_{\ell=1}^{N}\dfrac{D^{\rm a}_{\ell}|G_{1,\ell}|^2}{1+( \omega t^{\rm a}_{\ell})^{-2}}\bigg\}^2\label{eq-var}\\&~~~~~~~~~~~~~~~~~~~~~~~~~~~~~~~+8\gamma_{\rm L}\gamma_{\rm R}T_{\rm L}T_{\rm R}\omega^2|G_{1,N}|^2+8\gamma_{\rm L}T_{\rm L}\sum_{\ell=1}^{N}\dfrac{ D^{\rm a}_{\ell}|G_{1,\ell}|^2}{1+(\omega t^{\rm a}_{\ell})^{-2}}\bigg],\notag
\end{align}\end{widetext} 
where the first integral in Eq.~\eqref{eq-cur} corresponds to the heat current observed in Ref.~\cite{ht-2} for a harmonic chain when the particles have no activity. Similarly, the limit $D_{\ell}^{a}\to 0$ yields the second cumulant as shown in Ref.~\cite{ht-2}. An alternative derivation for first and second cumulants shown in Appendix \ref{JL-diff} gives the same cumulants as in Eqs.~\eqref{eq-cur} and \eqref{eq-var}. {\color{black}Similarly we use the scaled cumulant-generating function $\mu_{\rm R}(\lambda)$~\eqref{full-mu-R} to obtain the first and second scaled cumulants of the right heat flow $Q_{\rm R}$:
\begin{widetext}
\begin{align}
J_{\rm R}&=-\dfrac{T_{\rm L}-T_{\rm R}}{4\pi}\int_{-\infty}^{+\infty}~d\omega~4\gamma_{\rm L}\gamma_{\rm R} \omega^2 |G_{1,N}|^2-\dfrac{1}{4\pi}\int_{-\infty}^{+\infty}~d\omega~ 4\gamma_{\rm R}\sum_{\ell=1}^{N}\dfrac{D^{\rm a}_{\ell}|G_{N,\ell}|^2}{1+(\omega t^{\rm a}_{\ell})^{-2}},\label{eq-cur-R}\\
\dfrac{1}{\tau}[\langle Q_{\rm R}^2 \rangle-\langle Q_{\rm R} \rangle^2]&=\dfrac{1}{4\pi} \int_{-\infty}^{+\infty}~d\omega~\bigg[\bigg\{4\gamma_{\rm L}\gamma_{\rm R}(T_{\rm L}-T_{\rm R})\omega^2|G_{1,N}|^2+4\gamma_{\rm R}\sum_{\ell=1}^{N}\dfrac{D^{\rm a}_{\ell}|G_{N,\ell}|^2}{1+( \omega t^{\rm a}_{\ell})^{-2}}\bigg\}^2\label{eq-var-R}\\&~~~~~~~~~~~~~~~~~~~~~~~~~~~~~~~+8\gamma_{\rm L}\gamma_{\rm R}T_{\rm L}T_{\rm R}\omega^2|G_{1,N}|^2+8\gamma_{\rm R}T_{\rm R}\sum_{\ell=1}^{N}\dfrac{ D^{\rm a}_{\ell}|G_{N,\ell}|^2}{1+(\omega t^{\rm a}_{\ell})^{-2}}\bigg].\notag
\end{align}\end{widetext} 
}

Figures~\ref{cgf-fig}(a) and (b) show {\color{black}increasing observation time increases the} agreement of analytical expressions for the first two scaled cumulants of {\color{black}left} heat flow for a two-AOUP chain with numerical simulations performed using Langevin dynamics.
Figure~\ref{cgf-fig}(c) shows that as the particle activities $D_{i}^{\rm a}$ increase, the current $J_{\rm L}$ decreases and eventually changes sign. This is because the active forces perform work on the particles, so these particles dissipate heat to the reservoir to maintain steady state. Therefore, there is a competition between two currents: the current due to thermal forces (first term of Eq.~\eqref{eq-cur}) and that due to active forces (second term of Eq.~\eqref{eq-cur}). Figure~\ref{cgf-fig}(d) shows that these active forces enhance heat fluctuations. In summary, with increasing AOUP activity the distribution of the {\color{black}left} heat flow shifts to lower mean and its width increases. 

{\color{black}Similarly, Figs.~\ref{cgf-fig-R}(a) and (b), respectively, compare the analytical expressions for the first and second scaled cumulants of right heat flow given in Eqs.~\eqref{eq-cur-R} and \eqref{eq-var-R} with numerical simulations performed using Langevin dynamics.
In this case (for $T_{\rm L}>T_{\rm R}$), the {\color{black}right} current remains negative (leaving the chain and entering the bath) and decreases with the particle activity {\color{black}[see Fig.~\ref{cgf-fig-R}(c)]}.
{\color{black}This current's sign can be physically understood as follows. There are two currents that enter into the right heat bath: the thermal current due to the temperature gradient (first term of Eq.~\eqref{eq-cur-R}) and the current due to the active forces (second term of Eq.~\eqref{eq-cur-R}). 
Each make a negative contribution to $J_{\rm R}$. 
(Notice that for $J_{\rm L}$, the thermal current has opposite sign and competes with the active-force current, so the left current's sign depends on the dominant contribution.) 
Similar to the {\color{black}left heat flow}, here also the active forces enhance the fluctuations of the right heat flow; therefore, as the particle activity increases, the distribution shifts toward lower mean and broadens {\color{black}[see Fig.~\ref{cgf-fig-R}(d)]}.}

Figure~\ref{fig:ratio} shows for the three-AOUP chain the analytical ratios of left and right heat currents [Eqs.~\eqref{eq-cur} and \eqref{eq-cur-R}] and of the heat-flow variances [Eqs.~\eqref{eq-var} and \eqref{eq-var-R}], as functions of the {\color{black}leftmost particle's} activity $D_1^{\rm a}$ at fixed $D_2^{\rm a}$ for three different values of $D_3^{\rm a}$.
These cumulants are clearly neither anti-symmetric nor symmetric.

To gain more insight, Fig.~\ref{fig:ratio-same-d} shows these ratios when all particles have identical activity ($D_i^{\rm a} = D^{\rm a}~\forall~i$), as a function of the activity strength $D^{\rm a}$.
As expected, in the limit $D^{\rm a}\to0$, $J_{\rm L}\to -J_{\rm R}$ and $[\langle Q_{\rm L}^2 \rangle -\langle Q_{\rm L} \rangle^2] \to [\langle Q_{\rm R}^2 \rangle -\langle Q_{\rm R} \rangle^2]$. 
In the opposite limit ($D^{\rm a}\to\infty$), these ratios saturate to particular values (see dashed lines) obtained from {\color{black}the dominating contributions in the limit $D^{\rm a}~\to~\infty$ of} analytical expressions~\eqref{eq-cur}, \eqref{eq-cur-R}, \eqref{eq-var}, and \eqref{eq-var-R}.  
This can be physically understood as follows. 
When particle activity is sufficiently high that in Eqs.~\eqref{eq-cur} and \eqref{eq-cur-R} the active-force current (second integral) dominates the thermal current (first integral),
a majority of the heat current is due to the active forces and flows toward both heat baths, giving a positive ratio of currents. Similarly, in this limit the heat-flow fluctuations are mostly due to the active forces {\color{black}and their ratio saturates to its limiting behavior (dashed horizontal line in Fig.~\ref{fig:ratio-same-d}(b)) {\color{black} obtained from the dominating contribution.
}} 
Even when each particle has distinct activity,
we expect {\color{black}the ratio of cumulants of left and right heat flow in the limit of large activity strength of the particles (i.e., $D^{\rm a}_i\to \infty~\forall~i$) 
to saturate
(similar to Fig.~\ref{fig:ratio-same-d}) as long as the integrals in Eqs.~\eqref{eq-cur}, \eqref{eq-cur-R}, \eqref{eq-var}, and \eqref{eq-var-R} containing $D_i^{a}$ are dominating. } 
}

\section{Summary}
\label{summ}
In this paper, we considered a harmonic chain of $N$ active Ornstein-Uhlenbeck particles. Each particle is influenced by a persistent stationary-state Ornstein-Uhlenbeck active force which has an exponential correlation in time. The chain ends are connected via different friction constants to two heat reservoirs of different temperatures. Due to the temperature difference, heat generally flows through the system. We computed the steady-state heat flow entering each end of the chain, in the long-time limit analytically obtaining the characteristic function for this heat flow. 
We demonstrated two examples where one can compute the characteristic function for non-Gaussian observables. Finally, we used the characteristic function to compute the scaled cumulants for the heat flow and observed the effect of the activity on the heat current and its fluctuations. In particular, we found the activity of particles produces heat flow out the left end, thereby counteracting the rightward heat flow at the leftmost particle in the absence of activity. At the same time, it also enhances the fluctuations of the heat flow. In brief, activity of the particles reduces the mean and broadens the distribution of the left heat flow.
 
The results presented in this paper are based on the framework of stochastic thermodynamics~\cite{seifert-1,sekimoto} and give us an understanding of steady-state thermal conduction in an active-matter harmonic chain. Recent research has shown that the first two cumulants for an arbitrary current are constrained by the thermodynamic uncertainty relation~\cite{tur-1}: fluctuations of the current are bounded by entropy production. Therefore, these two cumulants will be useful in the thermodynamic uncertainty relation~\cite{tur-1} to infer the dissipation of this active-matter system. It would also be interesting to see the effect of active run-and-tumble particles and active Brownian particles on these first two cumulants and the related thermodynamic uncertainty relation.  
 
We emphasize that the large-deviation function for a stochastic observable (such as the heat flow) is related to the cumulant-generating function through the Legendre transform~\cite{ldf} (see Sec.~\ref{FP-sec}), where tails of the distribution are identified by the cutoffs within which $\mu(\lambda)$ is a real function. (See Refs.~\cite{Fogedby,pep-1,pep-2,pep-3} for the computation of these cutoffs.) The prefactor $g(\lambda)$ also importantly affects the tails of the distributions when $g(\lambda)$ has singularities~\cite{lde-2,lde-5,ht-9,Visco,apal,apal-2}.  Given our framework, one can consider simple examples permitting analytical computation of $\mu(\lambda)$ and $g(\lambda)$, and carefully invert the characteristic function to obtain the probability density function using methods discussed in Refs.~\cite{lde-5,ht-9,apal,apal-2}. 

In this paper, we have considered a harmonic chain where only the ends are connected to thermal reservoirs of different temperatures. Extending our system of active particles to connect each particle to a different temperature~\cite{Lebowitz,Falasco} would be interesting. {\color{black} Additionally, departures from Fourier's law for a harmonic chain composed of pinned active particles ({\color{black}where each particle is 
additionally 
confined in 
its own distinct
potential)} and the role of boundary conditions and system size on the heat conduction are interesting topics for future investigation}~\cite{adhar-review}.

\begin{acknowledgments}
This research was supported by ``Excellence Project 2018'' of the Cariparo foundation from the University of Padova (D.G.), a Natural Sciences and Engineering Research Council of Canada (NSERC) Discovery Grant (D.A.S.), and a Tier-II Canada Research Chair (D.A.S.). The authors thank Lorenzo Caprini for suggesting useful references, Rituparno Mandal for useful discussions, {\color{black}and the anonymous reviewers for their valuable suggestions that have improved the manuscript's content and presentation.}
\end{acknowledgments}
\bigskip

\appendix
\begin{widetext}
{\color{black}\section{The Fokker-Planck equation}
Here we derive the Fokker-Planck equation for the conditional moment-generating function $Z(\lambda,U,\tau|U_0)$, where $U \equiv [x_1,x_2,\dots,x_N,v_1,v_2,\dots,v_N,f_1,f_2,\dots,f_N]^\top$. We first write the evolution equation for the conditional joint density function  $\rho(Q_{\rm L},U,\tau|U_0)$ \cite{van}:
\begin{align}
 \dfrac{\partial \rho}{\partial t} &= -\sum_{i=1}^N\bigg[\dfrac{\partial}{\partial x_i} \bigg(\dfrac{\langle \Delta x_i\rangle}{\Delta t} \rho\bigg)+\dfrac{\partial}{\partial v_i} \bigg(\dfrac{\langle \Delta v_i\rangle}{\Delta t} \rho\bigg)+\dfrac{\partial}{\partial f_i} \bigg(\dfrac{\langle \Delta f_i\rangle}{\Delta t} \rho\bigg)\bigg] \nonumber\\
 &\quad +\dfrac{1}{2}\sum_{i,j}\bigg[\dfrac{\partial^2}{\partial x_i\partial x_j} \bigg(\dfrac{\langle \Delta x_i \Delta x_j\rangle}{\Delta t} \rho\bigg) +\dfrac{\partial^2}{\partial v_i\partial v_j} \bigg(\dfrac{\langle \Delta v_i \Delta v_j\rangle}{\Delta t} \rho\bigg)+\dfrac{\partial^2}{\partial f_i\partial f_j} \bigg(\dfrac{\langle \Delta f_i \Delta f_j\rangle}{\Delta t} \rho\bigg)\bigg] \nonumber\\
 &\quad +\dfrac{1}{2}\sum_{i,j}\bigg[\dfrac{\partial^2}{\partial x_i\partial v_j} \bigg(\dfrac{\langle \Delta x_i \Delta v_j\rangle}{\Delta t} \rho\bigg) +\dfrac{\partial^2}{\partial x_i\partial f_j} \bigg(\dfrac{\langle \Delta x_i \Delta f_j\rangle}{\Delta t} \rho\bigg)+\dfrac{\partial^2}{\partial v_i\partial f_j} \bigg(\dfrac{\langle \Delta v_i \Delta f_j\rangle}{\Delta t} \rho\bigg)\bigg] \\
 &\quad - \dfrac{\partial}{\partial Q_{\rm L}} \bigg(\dfrac{\langle \Delta Q_{\rm L}\rangle}{\Delta t}\rho\bigg) + \dfrac{1}{2}\dfrac{\partial^2}{\partial Q_{\rm L}^2}\bigg(\dfrac{\langle \Delta Q^2_{\rm L} \rangle}{\Delta t} \rho\bigg)\nonumber\\
 &\quad +\sum_{i=1}^{N}\bigg[\dfrac{\partial^2}{\partial Q_{\rm L}\partial x_{i}}\bigg(\dfrac{\langle \Delta x_i \Delta Q_{\rm L}\rangle}{\Delta t} \rho\bigg)+\dfrac{\partial^2}{\partial Q_{\rm L}\partial v_{i}}\bigg(\dfrac{\langle \Delta v_i \Delta Q_{\rm L}\rangle}{\Delta t} \rho\bigg)+\dfrac{\partial^2}{\partial Q_{\rm L}\partial f_{i}}\bigg(\dfrac{\langle \Delta f_i \Delta Q_{\rm L}\rangle}{\Delta t} \rho\bigg)\bigg].\notag
\end{align}
To evaluate the right-hand side, we discretize the dynamical equations \eqref{eq1}--\eqref{eq3} and {\color{black}(following the Stratonovich rule)} \eqref{heat-eqn}, and compute the moments in the limit of vanishing time-increment ($\Delta t \to 0$). Substituting these moments, we find
\begin{align}
    \dfrac{\partial \rho}{\partial t} &= -\sum_{i=1}^N\bigg[v_i\dfrac{\partial \rho}{\partial x_i}+\dfrac{1}{m_i}\dfrac{\partial}{\partial v_i} \bigg(\bigg\{-\sum_{j=1}^N \Phi_{i,j}x_j+f_i-\gamma_{\rm L}v_{1}\delta_{i,1}-\gamma_{\rm R}v_{N}\delta_{i,N}\bigg\}\rho\bigg)-\dfrac{1}{t_i^{\rm a}}\dfrac{\partial}{\partial f_i}(f_i \rho)\bigg]\\
    &\quad +\dfrac{\gamma_{\rm L}T_{\rm L}}{m_1^2}\dfrac{\partial^2 \rho}{\partial v_1^2}+\dfrac{\gamma_{\rm R}T_{\rm R}}{m_N^2}\dfrac{\partial^2 \rho}{\partial v_N^2}+\sum_{i=1}^{N} D_{i}^{\rm a} \dfrac{\partial^2 \rho}{\partial f_i^2}- \bigg(\dfrac{\gamma_{\rm L}T_{\rm L}}{m_1}-\gamma_{\rm L}v_1^2\bigg)\dfrac{\partial \rho}{\partial Q_{\rm L}} + \gamma_{\rm L}T_{\rm L}v_1^2\dfrac{\partial^2 \rho}{\partial Q_{\rm L}^2}+\dfrac{2\gamma_{\rm L}T_{\rm L}}{m_1} \dfrac{\partial^2}{\partial Q_{\rm L} v_1} (v_1 \rho).\notag
\end{align}
Fourier transforming $\rho$ to $Z(\lambda, U,\tau|U_0) \equiv \int_{-\infty}^{+\infty}~dQ_{\rm L}~\rho(Q_{\rm L},U,\tau|U_0)~e^{-\lambda Q_{\rm L}}$ gives the evolution equation~\eqref{fp-eqn-1} for the conditional moment-generating function {\color{black}$Z(\lambda, U,\tau|U_0)$}, in which the Fokker-Planck operator is
\begin{align}
    \mathcal{L}_\lambda &\equiv  -\sum_{i=1}^N\bigg[v_i\dfrac{\partial}{\partial x_i}+\dfrac{1}{m_i}\dfrac{\partial}{\partial v_i} \bigg(-\sum_{j=1}^N \Phi_{i,j}x_j+f_i-\gamma_{\rm L}v_{1}\delta_{i,1}-\gamma_{\rm R}v_{N}\delta_{i,N}\bigg)-\dfrac{1}{t_i^{\rm a}}\dfrac{\partial}{\partial f_i}f_i\bigg]\label{FPO}\\&+\dfrac{\gamma_{\rm L}T_{\rm L}}{m_1^2}\dfrac{\partial^2}{\partial v_1^2}+\dfrac{\gamma_{\rm R}T_{\rm R}}{m_N^2}\dfrac{\partial^2}{\partial v_N^2}+\sum_{i=1}^{N} D_{i}^{\rm a} \dfrac{\partial^2 }{\partial f_i^2}-\lambda \bigg(\dfrac{\gamma_{\rm L}T_{\rm L}}{m_1}-\gamma_{\rm L}v_1^2\bigg) + \lambda^2\gamma_{\rm L}T_{\rm L}v_1^2+\dfrac{2\lambda\gamma_{\rm L}T_{\rm L}}{m_1} \dfrac{\partial}{\partial v_1}v_1. \notag
\end{align}
}

\section{Detailed derivation of steady-state distribution $P_{\rm SS}(U)$}
\label{ss-app}
Here we compute the steady-state distribution (see Eq.~\eqref{ss-eqn}) for the full dynamics given in Eqs.~\eqref{eq1}--\eqref{eq3}. Fourier transforming Eqs.~\eqref{eq1}--\eqref{eq3} using Eq.~\eqref{FT-1} gives 
\begin{subequations}
\begin{align}
i\omega_n \tilde X(\omega_n)&=\tilde{V}(\omega_n)-\dfrac{\Delta X}{\tau},\label{X-eqn}\\
(i\omega_n M+\Gamma)\tilde V(\omega_n)&=-\Phi \tilde{X}(\omega_n)+\tilde F(\omega_n)+\tilde B(\omega_n)-\dfrac{M\Delta V}{\tau},\label{V-eqn}\\
\big(i\omega_n I+R^{-1}\big)\tilde F(\omega_n)&=\tilde{\mathcal{Z}} (\omega_n)-\dfrac{\Delta F}{\tau},\label{F-eqn}
\end{align}
\end{subequations}
where $[\Delta X,\Delta V,\Delta F] \equiv [X(\tau)-X(0), V(\tau)-V(0),F(\tau)-F(0)]${\color{black}. Notice that the phase space of our system is unbounded; therefore, these boundary terms cannot be neglected even in the long-time limit. 
In Eq.~\eqref{F-eqn}, $I$ is the identity matrix. }

Substituting Eqs.~\eqref{X-eqn} and \eqref{F-eqn} in Eq.~\eqref{V-eqn} gives
\begin{align}
(i\omega_n M+\Gamma)\tilde V(\omega_n)&=-\dfrac{\Phi}{i\omega_n} \bigg[\tilde{V}(\omega_n)-\dfrac{\Delta X}{\tau}\bigg]+\big[i\omega_n I+R^{-1}\big]^{-1}\bigg[\tilde {\mathcal{Z}}(\omega_n)-\dfrac{\Delta F}{\tau}\bigg]
+\tilde B(\omega_n)-\dfrac{M\Delta V}{\tau},
\end{align}
or, solving for $\tilde{V}$:
\begin{align}
\tilde V(\omega_n) &= i\omega_n G(\omega_n)\bigg[\big[i\omega_n I+R^{-1}\big]^{-1}\tilde{\mathcal{Z}}(\omega_n)+\tilde B(\omega_n)\bigg]+\dfrac{G(\omega_n)}{\tau}\bigg[\Phi\Delta X-i\omega_n M\Delta V -i\omega_n\big[i\omega_n I+R^{-1}\big]^{-1}\Delta F\bigg],  \label{Eq-FV}
\end{align}
where $G(\omega_n)\equiv[\Phi-\omega_n^2 M+i\omega_n\Gamma]^{-1}$ is the Green's function matrix. 

Using Eq.~\eqref{F-eqn}, the force vector at time $\tau$ {\color{black} can be computed using the inverse Fourier transform~\eqref{FT-2} and substituting $t=\tau-\epsilon$ (for $\epsilon > 0$), giving~\cite{ht-1, ht-2,apal-2}} 
\begin{subequations}
\begin{align}
F(\tau)&={\color{black}\lim_{\epsilon\to 0}\sum_{n=-\infty}^{+\infty} \tilde{F}(\omega_n)~e^{i\omega_n (\tau-\epsilon)}} \\
&=\lim_{\epsilon\to 0}\sum_{n=-\infty}^{+\infty} \tilde{F}(\omega_n)~e^{-i\omega_n \epsilon}\label{ftau}
\end{align}
\end{subequations}
{\color{black}for $\tau = 2\pi n/\omega_n$.}
In the limit of large $\tau$, we can convert the second summation into an integral over $\omega$. The matrix $\big[i\omega_n I+R^{-1}\big]^{-1}$ has only diagonal entries, and each entry gives a pole $\omega=i/t^{\rm a}_{\ell}$ which lies in the upper half of the complex $\omega$-plane. Therefore, using the Cauchy residue theorem~\cite{mmeth}, the second term {\color{black}(containing boundary terms)} vanishes in that limit, giving
\begin{align}
F(\tau)&=\lim_{\epsilon\to 0}\sum_{n=-\infty}^{+\infty}~e^{-i\omega_n  \epsilon}\big[i\omega_n I+R^{-1}\big]^{-1}\tilde{\mathcal{Z}}(\omega_n).
\end{align}

Similar to Eq.~\eqref{ftau}, inverse Fourier transforming Eq.~\eqref{Eq-FV} gives $V(\tau)$, and we find that the term 
\begin{align}
\sum_{n=-\infty}^{+\infty}e^{-i\omega_n \epsilon} \dfrac{G(\omega_n)}{\tau}\bigg[\Phi\Delta X-i\omega_nM\Delta V-i\omega_n\big[i\omega_n I+R^{-1}\big]^{-1}\Delta F\bigg],
\end{align}
vanishes in the $\tau\to\infty$ limit since all the poles lie in the upper half of the complex $\omega$-plane. Thus, the velocity vector at time $\tau$ is 
\begin{align}
V(\tau)=\lim_{\epsilon\to 0}\sum_{n=-\infty}^{+\infty}e^{-i\omega_n \epsilon}~i\omega_n G(\omega_n)\bigg[\big[i\omega_n I+R^{-1}\big]^{-1}\tilde{\mathcal{Z}}(\omega_n)+\tilde B(\omega_n)\bigg].
\end{align}
Substituting Eq.~\eqref{Eq-FV} in Eq.~\eqref{X-eqn} gives the Fourier-space position vector
\begin{align}
\tilde X(\omega_n)&=G(\omega_n)\bigg[\big[i\omega_n I+R^{-1}\big]^{-1}\tilde{\mathcal{Z}}(\omega_n)+\tilde B(\omega_n)\bigg]\\
&\quad +\dfrac{G(\omega_n)}{i\omega_n\tau}\bigg[\Phi\Delta X-i\omega_nM\Delta V-i\omega_n\big[i\omega_n I+R^{-1}\big]^{-1}\Delta F\bigg]-\dfrac{\Delta X}{i\omega_n\tau}. \nonumber
\end{align}

Following the same argument, in the limit of large $\tau$ the position vector at time $\tau$ is
\begin{align}
X(\tau)=\lim_{\epsilon\to 0}\sum_{n=-\infty}^{+\infty}e^{-i\omega_n \epsilon}~G(\omega_n)\bigg[\big[i\omega_n I+R^{-1}\big]^{-1}\tilde{\mathcal{Z}}(\omega_n)+\tilde B(\omega_n)\bigg].
\end{align}

Therefore, considering the contribution from $X(\tau)$, $V(\tau)$, and $F(\tau)$, the full state vector is
\begin{subequations}
\begin{align}
U^\top(\tau)&=\lim_{\epsilon\to 0}\sum_{n=-\infty}^{+\infty}e^{-i\omega_n \epsilon} \bigg(\bigg[\tilde {\mathcal{Z}}^\top(\omega_n)\big[i\omega_n I+R^{-1}\big]^{-1}+\tilde B^\top(\omega_n)\bigg]G^\top(\omega_n)\ , \label{Utop-tau} \\
&\quad\quad i\omega_n\bigg[\tilde {\mathcal{Z}}^\top(\omega_n)\big[i\omega_n I+R^{-1}\big]^{-1} +\tilde B^\top(\omega_n)\bigg]G^\top(\omega_n)\ ,\ \tilde {\mathcal{Z}}^\top(\omega_n)\big[i\omega_n I+R^{-1}\big]^{-1} \bigg),\nonumber\\
&=\lim_{\epsilon\to 0}\sum_{n=-\infty}^{+\infty}e^{-i\omega_n \epsilon}~\bigg[\sum_{j=1}^{N}\dfrac{\tilde \zeta_j(\omega_n)q_j^\top}{i\omega_n+1/t_{j}^{\rm a}}+\tilde{\eta}_{\rm L}(\omega_n)\ell_1^\top+\tilde{\eta}_{\rm R}(\omega_n)\ell_N^\top\bigg],\label{utop-vec}
\end{align}
\end{subequations}
where (for $j=1,\ldots,N$)
\begin{subequations}
\begin{align}
q_j^\top&
\equiv
(G_{1,j},G_{2,j},\dots,G_{N,j},i\omega_nG_{1,j},i\omega_nG_{2,j},\dots,i\omega_nG_{N,j},\delta_{1,j}\delta_{2,j},\dots,\delta_{N,j}),\\
\ell_j^\top &\equiv (G_{1,j},G_{2,j},\dots,G_{N,j},i\omega_nG_{1,j},i\omega_nG_{2,j},\dots,i\omega_nG_{N,j}, \overbrace{0,0,\dots,0}^{N}).
\end{align}
\end{subequations}
The first $N$ components correspond to positions, the next $N$ to velocities, and the last $N$ to OU forces. 

$U(\tau)$ has average zero and correlation
\begin{subequations}
\begin{align}
\langle U(\tau)U^\top(\tau) \rangle&=\lim_{\epsilon\to 0}\sum_{n=-\infty}^{+\infty}\sum_{m=-\infty}^{+\infty}e^{-i\omega_n \epsilon}~e^{-i\omega_m \epsilon}\nonumber\\
&\quad \times \bigg[\sum_{j=1}^{N}\sum_{p=1}^{N}\dfrac{q_j(\omega_n)\langle \tilde \zeta_j(\omega_n)\tilde \zeta_p(\omega_m)\rangle q_p^\top(\omega_m)}{(i\omega_n+1/t_{j}^{\rm a})(i\omega_m+1/t_{p}^{\rm a})}\\
&\quad\quad +\ell_1(\omega_n)\langle \tilde{\eta}_{\rm L}(\omega_n)\tilde{\eta}_{\rm L}(\omega_m)\rangle \ell_1^\top(\omega_m)+\ell_N(\omega_n) \langle \tilde{\eta}_{\rm R}(\omega_n)\tilde{\eta}_{\rm R}(\omega_m)\rangle \ell_N^\top(\omega_m)\bigg]\nonumber\\
&=\dfrac{2}{\tau}\sum_{n=-\infty}^{+\infty}\bigg[\sum_{j=1}^{N}\dfrac{D^{\rm a}_{j}q_jq_j^\dagger}{\omega^2_n+1/(t^{\rm a}_{j})^{2}} +\gamma_{\rm L}T_{\rm L}\ell_1\ell_1^\dagger+\gamma_{\rm R}T_{\rm R}\ell_N\ell_N^\dagger\bigg]\\
&=\dfrac{1}{\pi}\int_{-\infty}^{+\infty}~d\omega~\bigg[\sum_{j=1}^{N}\dfrac{D^{\rm a}_{j}q_jq_j^\dagger}{\omega^2+1/(t^{\rm a}_{j})^{2}}+\gamma_{\rm L}T_{\rm L}\ell_1\ell_1^\dagger+\gamma_{\rm R}T_{\rm R}\ell_N\ell_N^\dagger\bigg],\label{corr-eqn}
\end{align}
\end{subequations}
for Fourier-space noise correlations
\begin{subequations}\begin{align}
[\langle \tilde\eta_i(\omega)\tilde\eta_j(\omega') \rangle,\langle \tilde \zeta_i(\omega)\tilde \zeta_j(\omega') \rangle]^\top&=\dfrac{1}{\tau^2}\int_0^\tau~dt_1~\int_0^\tau~dt_2~e^{-i\omega t_1}~e^{-i\omega' t_2} [\langle \eta_i(t_1)\eta_j(t_2) \rangle,\langle \zeta_i(t_1)\zeta_j(t_2) \rangle]^\top\\
&=\frac{2}{\tau}\delta_{i,j}\delta_{\omega,-\omega'}[\gamma_iT_i,D_{i}^{\rm a}]^\top. \label{noise-correlations}
\end{align}
\end{subequations}
A Gaussian with zero mean and correlation~\eqref{corr-eqn} gives the reported steady-state distribution~\eqref{ss-eqn}.

\section{Detailed derivation of Eq.~\eqref{z-u-u0}}
\label{dd-eq}
Here we present the detailed derivation of Eq.~\eqref{z-u-u0}, helpful for computing the characteristic function for the heat $Q_{\rm L}$ entering the leftmost particle from the left bath. Fourier transforming~\eqref{FT-2} the RHS of $Q_{\rm L}$ in Eq.~\eqref{heat-eqn} gives
\begin{align}
Q_{\rm L}=\dfrac{\tau}{2}\sum_{n=-\infty}^{+\infty}~[\tilde{\eta}_{\rm L}(\omega_n) \tilde v_1(-\omega_n)+\tilde{\eta}_{\rm L}(-\omega_n) \tilde v_1(\omega_n)-2\gamma_{\rm L} \tilde v_1 (\omega_n) \tilde v_1(-\omega_n)].\label{Q-L-omega}
\end{align}

Using Eq.~\eqref{Eq-FV}, we can write the Fourier-space velocity of the leftmost particle as
\begin{align}
\tilde v_1(\omega_n)&=i\omega_n\bigg[\sum_{\ell=1}^{N} \dfrac{G_{1,\ell}(\omega_n)\tilde{\zeta}_\ell(\omega_n)}{i\omega_n+1/t^{\rm a}_{\ell}}+G_{1,1}(\omega_n)\tilde{\eta}_{\rm L}(\omega_n)+G_{1,N}(\omega_n)\tilde{\eta}_{\rm R}(\omega_n)\bigg]+\dfrac{\mathcal{R}^\top \Delta U}{\tau},\label{v1-eqn}
\end{align}
for
\begin{subequations}
\begin{align} 
\mathcal{R}^\top&\equiv \bigg[[G(\omega_n)\Phi]_{1,1}\ ,\ [G(\omega_n)\Phi]_{1,2}\ ,\dots,\ [G(\omega_n)\Phi]_{1,N}\ ,\ -i\omega_n [G(\omega_n)M]_{1,1},\label{q-eqn-1}\\
&\quad -i\omega_n [G(\omega_n)M]_{1,2},\dots,-i\omega_n [G(\omega_n)M]_{1,N},
-i\omega_n\bigg(\frac{G_{1,1}(\omega_n)}{i\omega_n+1/t_{1}^{\rm a}},\frac{G_{1,2}(\omega_n)}{i\omega_n+1/t_{2}^{\rm a}},\dots,\frac{G_{1,N}(\omega_n)}{i\omega_n+1/t^{\rm a}_{N}}\bigg)\bigg],\nonumber\\
\Delta U&{\color{black}\equiv}[\Delta x_1,\Delta x_2,\dots,\Delta x_N,\Delta v_1,\Delta v_2,\dots,\Delta v_N,\Delta f_1,\Delta f_2,\dots,\Delta f_N]^\top.
\end{align}
\end{subequations}

Similarly, replacing $\omega_n$ in Eq.~\eqref{v1-eqn} by $-\omega_n$,
\begin{align}
\tilde v_1(-\omega_n)&=-i\omega_n\bigg[ \sum_{\ell=1}^{N} \dfrac{G_{1,\ell}(-\omega_n)\tilde{\zeta}_\ell(-\omega_n)}{-i\omega_n+1/t^{\rm a}_{\ell}}+G_{1,1}(-\omega_n)\tilde{\eta}_{\rm L}(-\omega_n)+G_{1,N}(-\omega_n)\tilde{\eta}_{\rm R}(-\omega_n)\bigg]+\dfrac{\mathcal{R}^\dagger \Delta U}{\tau}.\label{v1m-eqn}
\end{align}

Substituting Eq.~\eqref{v1-eqn} and \eqref{v1m-eqn} in Eq.~\eqref{Q-L-omega} gives
\begin{align}
Q_{\rm L}=\dfrac{1}{2}\sum_{n=-\infty}^{+\infty}\bigg[\tau\tilde \xi_n^\top C_n \xi_n^*+ \tilde\xi_n^\top \alpha_n+ \alpha_{-n}^\top \tilde \xi_n^*-2\gamma_{\rm L}\dfrac{\Delta U^\top \mathcal{R}\mathcal{R}^\dagger \Delta U}{\tau}\bigg],\label{Q-eqn-FT}
\end{align}
{\color{black}where $\tilde \xi_{n}^\top \equiv[\tilde \zeta_1(\omega_n), \dots, \tilde \zeta_N(\omega_n), \tilde \eta_{\rm L}(\omega_n),\tilde \eta_{\rm R}(\omega_n)]$ is the Fourier-space noise vector, and the Hermitian matrix $C_n$ and vectors $\alpha_{-n}^\top$ are, respectively,}
\begin{subequations}
\begin{align}
C_n&\equiv C_n^{(1)}-2\gamma_{\rm L} C_{n}^{(2)},\label{cn-eqn}\\
\alpha_{-n}^\top&\equiv A_{-n}^\top-2\gamma_{\rm L} B_{-n}^\top.\label{alpha-eqn}
\end{align}
\end{subequations}
In Eq.~\eqref{cn-eqn}, the upper-triangle matrix elements (for $1\leq i\leq j \leq N$) of Hermitian matrices $C_n^{(1)}$ and $C_n^{(2)}$, respectively, are
\begin{subequations}
\begin{align}
[C^{(1)}_{n}]_{i,j}&=0{\color{black}\ ,}\\
[C^{(1)}_{n}]_{i,N+1}&=\dfrac{G_{1,i}}{1+(i\omega_n t_{i}^{\rm a})^{-1}}{\color{black}\ ,}\\ 
[C^{(1)}_{n}]_{i,N+2}&=0{\color{black}\ ,}\\
[C^{(1)}_{n}]_{N+1,N+1}&=i\omega_n[G_{1,1}-G_{1,1}^*]{\color{black}\ ,}\\
[C^{(1)}_{n}]_{N+1,N+2}&=-i\omega_n G_{1,N}^*{\color{black}\ ,}\\
[C^{(1)}_{n}]_{N+2,N+2}&=0{\color{black}\ ,}
\end{align}
\end{subequations}
and\begin{subequations}
\begin{align}
[C_n^{(2)}]_{i,j}&=\dfrac{\omega_n^2 G_{1,i}G_{1,j}^*}{(i\omega_n+1/t^{\rm a}_{i})(-i\omega_n+1/t^{\rm a}_{j})}{\color{black}\ ,}\\
[C_n^{(2)}]_{i,N+1}&=\dfrac{\omega_n^2G_{1,i}G_{1,1}^*}{i\omega_n+1/t_{i}^{\rm a}}{\color{black}\ ,} \\
[C_n^{(2)}]_{i,N+2}&=\dfrac{\omega_n^2 G_{1,i}G_{1,N}^*}{i\omega_n+1/t_{i}^{\rm a}}{\color{black}\ ,}\\
[C^{(2)}_{n}]_{N+1,N+1}&=\omega_n^2G_{1,1}G_{1,1}^*{\color{black}\ ,}\\
[C^{(2)}_{n}]_{N+1,N+2}&=\omega_n^2G_{1,1} G_{1,N}^*{\color{black}\ ,}\\
[C^{(2)}_{n}]_{N+2,N+2}&=\omega_n^2G_{1,N} G_{1,N}^*\ .
\end{align}\end{subequations}
Further, the two vectors in $\alpha_{-n}^\top$ in Eq.~\eqref{alpha-eqn} are
\begin{subequations}
\begin{align}
A_{-n}^\top&\equiv (\underbrace{0,0,\dots,0}_{N},1,0)\Delta U^\top \mathcal{R},\\
B_{-n}^\top&\equiv \bigg(\frac{-G_{1,1}^*}{-1+(i\omega_n t_{1}^{\rm a})^{-1}},\frac{-G_{1,2}^*}{-1+(i\omega_n t_{2}^{\rm a})^{-1}},\dots,\frac{-G_{1,N}^*}{-1+(i\omega_n t_{N}^{\rm a})^{-1}},-i\omega_n G_{1,1}^*,-i\omega_n G_{1,N}^*\bigg) \Delta U^\top \mathcal{R}.
\end{align}
\end{subequations}

We now compute the conditional characteristic function for $Q_{\rm L}$ defined as (see Eq.~\eqref{eq-z})
\begin{subequations}
\begin{align}
Z(\lambda,U,\tau|U_0)&\equiv \bigg\langle e^{-\lambda Q_{\rm L}} \delta[U-U(\tau)]\bigg\rangle_{U_0}\\
&=\int \dfrac{d^{3N}\sigma}{(2\pi)^{3N}}e^{i\sigma^\top U} \big\langle e^{E(\tau)}\big\rangle_{U_0},\label{RCF}
\end{align}
\end{subequations}
where $E(\tau)\equiv -\lambda Q_{\rm L}-i U^\top(\tau) \sigma$. The second line results from substituting the integral representation of the Dirac delta function. The state vector $U(\tau)$ can be rewritten using Eq.~\eqref{utop-vec} as
\begin{subequations}
\begin{align}
U^\top(\tau)&=\lim_{\epsilon\to 0}\sum_{n=-\infty}^{+\infty}e^{-i\omega_n \epsilon}\left[\sum_{j=1}^N\dfrac{\tilde{\zeta}_j(\omega_n)q_j^\top}{i\omega_n+1/t_{j}^{\rm a}}+\tilde{\eta}_{\rm L}(\omega_n)\ell_1^\top+\tilde{\eta}_{\rm R}(\omega_n)\ell_N^\top \right]\\
&\equiv\lim_{\epsilon\to 0}\sum_{n=-\infty}^{+\infty}e^{-i\omega_n \epsilon}~\tilde \xi_n^\top K_1,\label{k1-eqn}\\
U(\tau)&= \lim_{\epsilon\to 0}\sum_{n=-\infty}^{+\infty}e^{-i\omega_n \epsilon}\left[\sum_{j=1}^N~\dfrac{q_j\tilde{\zeta}_j(\omega_n)}{i\omega_n+1/t_{j}^{\rm a}}+\ell_1\tilde{\eta}_{\rm L}(\omega_n)+\ell_N\tilde{\eta}_{\rm R}(\omega_n) \right]\\
&\equiv\lim_{\epsilon\to 0}\sum_{n=-\infty}^{+\infty}e^{-i\omega_n \epsilon}~K_2^\top~\tilde \xi_n\label{k2-eqn},
\end{align}
\end{subequations}
where the inner products in \eqref{k1-eqn} and \eqref{k2-eqn} are defined using the respective column and row vectors, 
\begin{subequations}
\begin{align}  
& ~~~~~~~~~~~~~~~~~~~~~~~~~~~K_1\equiv \begin{bmatrix}
    (i\omega_n+1/t_{1}^{\rm a})^{-1}q_1^\top\\
    (i\omega_n+1/t_{2}^{\rm a})^{-1}q_2^\top\\
    \vdots\\
    (i\omega_n+1/t^{\rm a}_{N})^{-1}q_N^\top\\
    \ell_1^\top\\
    \ell_N^\top
    \end{bmatrix},\\
K_2^\top&\equiv [(i\omega_n+1/t_{1}^{\rm a})^{-1}q_1,(i\omega_n+1/t_{2}^{\rm a})^{-1}q_2,\dots,(i\omega_n+1/t^{\rm a}_{N})^{-1}q_N,\ell_1,\ell_N],
\end{align}
\end{subequations}
in which the first $N$ components correspond to active noise and the last two to thermal noise. Substituting $U^\top(\tau)$ in $E(\tau)$ leads to
\begin{align}
E(\tau)&=\sum_{n=1}^{+\infty}\bigg[-\lambda \tau\tilde \xi_n^\top C_n \tilde \xi_n^*+\tilde\xi_n^\top \beta_n+ \beta_{-n}^\top \tilde \xi_n^*+\dfrac{2\gamma_{\rm L}\lambda}{\tau}|\Delta_n|^2\bigg]-\dfrac{1}{2}\lambda \tau\tilde \xi_0^\top C_0 \tilde\xi_0+\tilde\xi_0^\top \beta_0+\dfrac{\lambda \gamma_{\rm L}}{\tau} |\Delta_0|^2,\label{e-tau}
\end{align}
for\begin{subequations}
\begin{align}
|\Delta_n|^2&\equiv \Delta U^\top \mathcal{R}\mathcal{R}^\dagger \Delta U,\\
\beta_n&\equiv -\lambda \alpha_n-i e^{-i\omega_n \epsilon} \big(K_1\sigma\big).
\end{align}
\end{subequations}

The average appearing in Eq.~\eqref{RCF}, {\color{black}$\left\langle e^{E(\tau)} \right\rangle_{U_0}$, can be computed as follows. We first note that $E(\tau)$ (given in Eq.~\eqref{e-tau}) is quadratic in $\tilde \xi_n$ and in $\tilde \xi_0$. Since $\tilde \xi_0$ and each $\tilde \xi_n$ ($n=1,2,\dots, \infty$) are independent and identically distributed noise vectors,
\begin{subequations}
\begin{align}
P(\tilde \xi_n)&=\dfrac{e^{-\tilde \xi_n^\top \Lambda^{-1}\tilde \xi_n^*}}{\pi^{N+2}\det[\Lambda]}\ , \quad n\geq 1  \label{n-d-1},\\
P(\tilde \xi_0)&=\dfrac{e^{-\frac{1}{2}\tilde \xi_0^\top \Lambda^{-1}\tilde \xi_0}}{\sqrt{(2\pi)^{N+2}\det[\Lambda]}}.\label{n-d-0} \ ,
\end{align}
\end{subequations}
we write $\big\langle e^{E(\tau)}\big\rangle_{U_0}$ in product form:}
\begin{align}
\langle e^{E(\tau)}\rangle_{U_0}&=\prod_{n=1}^{\infty}\bigg\langle e^{-\lambda \tau\tilde \xi_n^\top C_n \tilde \xi_n^*+\tilde\xi_n^\top \beta_n+ \beta_{-n}^\top \tilde \xi_n^*+\frac{2\gamma_{\rm L}\lambda}{\tau}|\Delta_n|^2}\bigg\rangle_{U_0}  \bigg\langle  e^{-\frac{1}{2}\lambda \tau\tilde \xi_0^\top C_0 \tilde\xi_0+\tilde\xi_0^\top \beta_0+\frac{\lambda \gamma_{\rm L}}{\tau} |\Delta_0|^2} \bigg\rangle_{U_0}. \label{this-eqn}
\end{align}
In \eqref{n-d-1} and \eqref{n-d-0}, the diagonal matrix $\Lambda\equiv 2/\tau~\text{diag}(D^{\rm a}_{1},D^{\rm a}_{2},\dots,D^{\rm a}_{N},\gamma_{\rm L} T_{\rm L},\gamma_{\rm R} T_{\rm R})$ carries information about the strength of thermal and active noises.

{\color{black}We compute each average by Gaussian integration, simplifying} Eq.~\eqref{this-eqn} to
\begin{align}
\langle e^{E(\tau)}\rangle_{U_0}&=\exp\bigg(-\dfrac{1}{2}\sum_{n=-\infty}^{+\infty} \ln\big[\det(\Lambda \Omega_n)\big]\bigg)\exp\bigg(\sum_{n=\infty}^{+\infty}\bigg[\dfrac{1}{2}\beta_{-n}^T\Omega_n^{-1}\beta_n+\frac{\lambda\gamma_{\rm L}}{\tau}|\Delta_n|^2\bigg]\bigg),\label{gf}
\end{align}
for $\Omega_n\equiv \Lambda^{-1}+\lambda\tau C_n$.

In the limit of large $\tau$, these summations become integrals, converting \eqref{gf} to 
\begin{align}
\langle e^{E(\tau)}\rangle_{U_0}&\approx e^{\tau \mu(\lambda)}e^{-\frac{1}{2}\sigma^\top H_1(\lambda)\sigma+i\Delta U^\top H_2(\lambda)\sigma+\frac{1}{2}\Delta U^\top H_3(\lambda)\Delta U}.\label{av-E}
\end{align}
The exponent $\mu(\lambda)$ in the integral form is 
{\color{black}
\begin{align}
\mu(\lambda)\equiv-\dfrac{1}{4\pi} \int_{-\infty}^{+\infty}~d\omega~\ln [\det(\Lambda \Omega)],\label{mu-lam-app}
\end{align} } 
and the matrices are\begin{subequations}
\begin{align}
H_1(\lambda)&\equiv\dfrac{\tau}{2\pi}\int_{-\infty}^{+\infty}d\omega~ K_2^\dagger \Omega^{-1}K_1,\label{h1-eqn}\\
H_2(\lambda)&\equiv\dfrac{\lambda\tau}{2\pi}\int_{-\infty}^{+\infty}d\omega~e^{-i\omega \epsilon}~a_1^\top \Omega^{-1} K_1,\label{h2-eqn}\\
H_3(\lambda)&\equiv\dfrac{\lambda\tau}{2\pi}\int_{-\infty}^{+\infty}d\omega~\bigg[\lambda a_1^\top \Omega^{-1}a_2+\frac{2\gamma_{\rm L}}{\tau} \mathcal{R}\mathcal{R}^\dagger\bigg],\label{h3-eqn}
\end{align}
\end{subequations}
for vectors
\begin{subequations}
\begin{align}
a_1^\top&\equiv\bigg(\frac{2\gamma_{\rm L}G_{1,1}^*\mathcal{R}}{-1+(i\omega_n t_{1}^{\rm a})^{-1}}\ ,\ \frac{2\gamma_{\rm L}G_{1,2}^* \mathcal{R}}{-1+(i\omega_n t_{2}^{\rm a})^{-1}}\ ,\dots,\ \frac{2\gamma_{\rm L}G_{1,N}^*\mathcal{R}}{-1+(i\omega_n t^{\rm a}_{N})^{-1}}\ , [1+2\gamma_{\rm L}i\omega_n G_{1,1}^*]\mathcal{R}\ ,\ 2\gamma_{\rm L}i\omega_n  G_{1,N}^* \mathcal{R}\bigg),\\
a_2&\equiv\bigg(\frac{-2\gamma_{\rm L}G_{1,1}\mathcal{R}^\dagger}{1+(i\omega_n t_{1}^{\rm a})^{-1}}\ ,\ \frac{-2\gamma_{\rm L}G_{1,2} \mathcal{R}^\dagger}{1+(i\omega_n t_{2}^{\rm a})^{-1}}\ ,\dots,\ \frac{-2\gamma_{\rm L}G_{1,N}\mathcal{R}^\dagger}{1+(i\omega_n t^{\rm a}_{N})^{-1}}\ ,
 [1-2\gamma_{\rm L}i\omega_n G_{1,1}]\mathcal{R}^\dagger\ ,\ -2\gamma_{\rm L}i\omega_n G_{1,N} \mathcal{R}^\dagger\bigg)^\top.
\end{align}
\end{subequations}
Note that in $a_1^\top$ and $a_2$, the first $N$ elements correspond to active noises and the last two to thermal noises.

{\color{black}
Substituting Eq.~\eqref{av-E} in Eq.~\eqref{RCF} and integrating over $\sigma$ yields
\begin{align}
&Z(\lambda,U,\tau|U_0)\approx e^{\tau \mu(\lambda)}\dfrac{e^{\frac{1}{2}\Delta U^\top H_3(\lambda)\Delta U}}{\sqrt{(2\pi)^{3N}\det H_1(\lambda)}} e^{-\frac{1}{2} [U^\top+\Delta U^\top H_2(\lambda)]~H_1^{-1}(\lambda)~[U+H_2^\top(\lambda)\Delta U]}.\label{this-eqn-1}  
\end{align}
The formal long-time solution of the Fokker-Planck equation (see Sec.~\ref{FP-sec}) is $Z(\lambda,U,\tau|U_0)\approx e^{\tau \mu(\lambda)}\chi(U_0,\lambda)\Psi(U,\lambda)$.
Therefore, to identify the left- and right-eigenfunctions, {\color{black}we factorize the RHS of Eq.~\eqref{this-eqn-1}} into separate factors that capture the respective dependence on $U$ and $U_0$. 
This identification can be achieved by setting $(H_1^{-1}H_2^\top-H_3+H_2 H_1^{-1}H_2^\top)+(H_2H_1^{-1}-H_3+H_2 H_1^{-1}H_2^\top)^\top=0$, giving \eqref{z-u-u0}.}

\section{Alternative derivation of first and second scaled cumulant for left heat flow}
\label{JL-diff}
Here we derive the first two scaled cumulants for the left heat flow, starting from its Fourier representation~\eqref{Q-L-omega}. This calculation verifies the cumulants obtained from the cumulant-generating function $\mu(\lambda)$ (see Eq.~\eqref{full-mu}). The computation of higher cumulants (above second) using the following method becomes complicated, and therefore it is convenient to compute the cumulants from $\mu(\lambda)$. 

We first obtain the first scaled cumulant. In \eqref{Q-L-omega}, we will substitute the Fourier-transformed velocity $\tilde v_1(\omega_n)$ of the first particle~\eqref{v1-eqn}. We first recall from Sec.~\ref{cum} that in the long-time limit the cumulant-generating function is independent of $g(\lambda)$, which generally captures the boundary contributions. Therefore, dropping the boundary contributions in this limit simplifies $\tilde v_{1}(\omega_n)$ to
\begin{align}
\tilde v_1(\omega_n)\approx i\omega_n\bigg[\sum_{\ell=1}^{N} \dfrac{G_{1,\ell}(\omega_n)}{i\omega_n+1/t^{\rm a}_{\ell}}\tilde{\zeta}_\ell(\omega_n)+G_{1,1}(\omega_n)\tilde{\eta}_{\rm L}(\omega_n)+G_{1,N}(\omega_n)\tilde{\eta}_{\rm R}(\omega_n)\bigg] \ .\label{v-eqn-sm}
\end{align}
Substituting this in Eq.~\eqref{Q-L-omega} and averaging over both thermal and active noise gives
\begin{align}
    \langle Q_{\rm L} \rangle &\approx\sum_{n=-\infty}^{+\infty} \bigg\{-i\omega_n [G_{1,1}(-\omega_n)-G_{1,1}(\omega_n)]\gamma_{\rm L} T_{\rm L} -2\gamma_{\rm L}\omega_n^2\bigg(\sum_{\ell=1}^{N}\dfrac{D^{\rm a}_{\ell} |G_{1,\ell}|^2}{\omega_n^2+1/(t^{\rm a}_{\ell})^{2}}
    +|G_{1,1}|^2 \gamma_{\rm L}T_{\rm L}+|G_{1,N}|^2 \gamma_{\rm R}T_{\rm R}\bigg)\bigg\} \ . \label{this-new-eq-1}
\end{align}
Using the definition of the Green's function matrix~\eqref{gr-fn-def},
\begin{align}
    G(-\omega_n)-G(\omega_n)=2i\omega_n G(-\omega_n)\Gamma G(\omega_n) \ ,
\end{align}
thus
\begin{subequations}
\begin{align}
    G_{1,1}(-\omega_n)-G_{1,1}(\omega_n)&=2i\omega_n \sum_{\ell,m} G_{1,\ell}(-\omega_n)\Gamma_{\ell,m} G_{m,1}(\omega_n)\\
    &=2i\omega_n \sum_{\ell,m} G_{1,\ell}(-\omega_n)\delta_{\ell,m}[\delta_{\ell,1}\gamma_{\rm L}+\delta_{m,N}\gamma_{\rm R}] G_{m,1}(\omega_n)\\
    &=2i\omega_n \sum_{\ell} G_{1,\ell}(-\omega_n)[\delta_{\ell,1}\gamma_{\rm L}+\delta_{\ell,N}\gamma_{\rm R}] G_{\ell,1}(\omega_n)\\
    &=2i\omega_n (|G_{1,1}|^2\gamma_{\rm L}+|G_{1,N}|^2\gamma_{\rm R}) \ , \label{c-8}
\end{align}
\end{subequations}
where the last line follows from the symmetry of the Green's function matrix \eqref{g-fun}. Substituting \eqref{c-8} in the first term inside curly brackets in \eqref{this-new-eq-1}, and converting the summation into a time integral in the long-time limit, gives $J_{\rm L}$ as in \eqref{eq-cur}.

Next, we compute the second scaled cumulant for the left heat flow. 
We square both sides of Eq.~\eqref{Q-L-omega} to write
\begin{subequations}
\begin{align}
Q_{\rm L}^2&= \dfrac{\tau^2}{4}\sum_{n,m=-\infty}^{+\infty}[\tilde{\eta}_{\rm L}(\omega_n) \tilde v_1(-\omega_n)+\tilde{\eta}_{\rm L}(-\omega_n) \tilde v_1(\omega_n)-2\gamma_{\rm L} \tilde v_1 (\omega_n) \tilde v_1(-\omega_n)]~~\times\\
&~~~~~~~~~~~~~~~~~~~~[\tilde{\eta}_{\rm L}(\omega_m) \tilde v_1(-\omega_m)+\tilde{\eta}_{\rm L}(-\omega_m) \tilde v_1(\omega_m)-2\gamma_{\rm L} \tilde v_1 (\omega_m) \tilde v_1(-\omega_m)]\nonumber\\
&=\dfrac{\tau^2}{4}\sum_{n,m=-\infty}^{+\infty}[\tilde{\eta}_{\rm L}(\omega_n) \tilde v_1(-\omega_n)\tilde{\eta}_{\rm L}(\omega_m) \tilde v_1(-\omega_m)+\tilde{\eta}_{\rm L}(-\omega_n) \tilde v_1(\omega_n)\tilde{\eta}_{\rm L}(-\omega_m) \tilde v_1(\omega_m)\\&~~~~~~~~~~~~~~~~~~-4 \gamma_{\rm L} \tilde{\eta}_{\rm L}(\omega_n) \tilde v_1(-\omega_n)\tilde v_1 (\omega_m) \tilde v_1(-\omega_m)-4 \gamma_{\rm L} \tilde{\eta}_{\rm L}(-\omega_n) \tilde v_1(\omega_n)\tilde v_1 (\omega_m) \tilde v_1(-\omega_m)\nonumber\\&~~~~~~~~~~~~~~~~~~+2\tilde{\eta}_{\rm L}(\omega_n) \tilde v_1(-\omega_n)\tilde{\eta}_{\rm L}(-\omega_m) \tilde v_1(\omega_m)+4 \gamma_{\rm L}^2\tilde v_1 (\omega_n) \tilde v_1(-\omega_n)\tilde v_1 (\omega_m) \tilde v_1(-\omega_m)] \ . \nonumber
\end{align}
\end{subequations}
Averaging over the noise distributions gives
\begin{align}
\langle Q_{\rm L}^2\rangle-\langle Q_{\rm L} \rangle^2&= \dfrac{\tau^2}{4}\sum_{n,m=-\infty}^{+\infty}[\langle \tilde{\eta}_{\rm L}(\omega_n) \tilde{\eta}_{\rm L}(\omega_m)  \rangle \langle \tilde v_1(-\omega_n)\tilde v_1(-\omega_m)\rangle+\langle\tilde{\eta}_{\rm L}(-\omega_n)  \tilde{\eta}_{\rm L}(-\omega_m) \rangle\langle \tilde v_1(\omega_n) \tilde v_1(\omega_m)\rangle\label{big-eqn-1}\\&
-4 \gamma_{\rm L} \langle \tilde{\eta}_{\rm L}(\omega_n) \tilde v_1 (\omega_m)  \rangle \langle \tilde v_1(-\omega_n)\tilde v_1(-\omega_m)\rangle-4 \gamma_{\rm L} \langle \tilde{\eta}_{\rm L}(-\omega_n) \tilde v_1 (\omega_m) \rangle\langle \tilde v_1(\omega_n)\tilde v_1(-\omega_m)\rangle\nonumber\\&
+2\langle\tilde{\eta}_{\rm L}(\omega_n) \tilde{\eta}_{\rm L}(-\omega_m)\rangle\langle \tilde v_1(-\omega_n) \tilde v_1(\omega_m)\rangle+4 \gamma_{\rm L}^2\langle\tilde v_1 (\omega_n) \tilde v_1 (\omega_m) \rangle\langle  \tilde v_1(-\omega_n)\tilde v_1(-\omega_m)\rangle +\nonumber\\&
+ \langle \tilde{\eta}_{\rm L}(\omega_n) \tilde v_1(-\omega_m)\rangle \langle \tilde{\eta}_{\rm L}(\omega_m) \tilde v_1(-\omega_n)\rangle+\langle\tilde{\eta}_{\rm L}(-\omega_n)  \tilde v_1(\omega_m) \rangle\langle \tilde{\eta}_{\rm L}(-\omega_m)\tilde v_1(\omega_n)\rangle\nonumber\\& -4 \gamma_{\rm L} \langle \tilde{\eta}_{\rm L}(\omega_n)  \tilde v_1(-\omega_m) \rangle \langle \tilde v_1 (\omega_m) \tilde v_1(-\omega_n)\rangle-4 \gamma_{\rm L} \langle \tilde{\eta}_{\rm L}(-\omega_n)  \tilde v_1(-\omega_m) \rangle\langle\tilde v_1 (\omega_m) \tilde v_1(\omega_n)\rangle\nonumber\\&+2\langle\tilde{\eta}_{\rm L}(\omega_n) \tilde v_1(\omega_m)\rangle\langle\tilde{\eta}_{\rm L}(-\omega_m) \tilde v_1(-\omega_n)\rangle+4 \gamma_{\rm L}^2\langle\tilde v_1 (\omega_n) \tilde v_1(-\omega_m) \rangle\langle\tilde v_1 (\omega_m) \tilde v_1(-\omega_n) \rangle      ], \nonumber
\end{align}
where we have used Wick's theorem~\cite{Kardar} for multivariate Gaussian distributions.
We substitute $\tilde v_1(\omega_n)$ from Eq.~\eqref{v-eqn-sm} on the right-hand side of \eqref{big-eqn-1}, then write the average over thermal and active noise in each term utilizing Eq.~\eqref{noise-correlations}.
This eventually leads to
\begin{align}
\dfrac{\langle Q_{\rm L}^2\rangle-\langle Q_{\rm L} \rangle^2}{\tau}&\approx \dfrac{1}{\tau}\sum_{n=-\infty}^{+\infty}\bigg[(1-4\omega_n^2\gamma_{\rm L}^2 |G_{1,1}|^2-4\omega_n^2\gamma_{\rm L}\gamma_{\rm R} |G_{1,N}|^2)~\times\\
&\quad \quad \bigg(\sum_{\ell=1}^{N}   \dfrac{4\gamma_{\rm L}T_{\rm L} \omega_n^2D_{\ell}^{\rm a}|G_{1,\ell}(\omega_n)|^2}{\omega_n^2+(t_\ell^{\rm a})^{-2}}+4\gamma_{\rm L}^2T_{\rm L}^2 \omega_n^2|G_{1,1}(\omega_n)|^2+4\gamma_{\rm L}T_{\rm L}\gamma_{\rm R}T_{\rm R} \omega_n^2|G_{1,N}(\omega_n)|^2\bigg)\nonumber\\
&\quad +2\bigg(\sum_{\ell=1}^{N}  \dfrac{2\gamma_{\rm L} \omega_n^2D_{\ell}^{\rm a} |G_{1,\ell}(\omega_n)|^2}{\omega_n^2+(t_\ell^{\rm a})^{-2}}+2\gamma_{\rm L}^2T_{\rm L}\omega_n^2|G_{1,1}(\omega_n)|^2+2\gamma_{\rm L}\gamma_{\rm R}T_{\rm R}\omega_n^2|G_{1,N}(\omega_n)|^2\bigg)^2 \nonumber\\
&\quad -2\omega_n^2\gamma_{\rm L}^2T_{\rm L}^2 \big\{[G_{1,1} (-\omega_n)]^2 + [G_{1,1} (\omega_n)]^2\big\}\bigg].\nonumber
\end{align}
In the long-time limit, the summation becomes a time integral, giving Eq.~\eqref{eq-var}. 
\end{widetext}


\begin{thebibliography}{2}%
\makeatletter
\providecommand \@ifxundefined [1]{%
 \@ifx{#1\undefined}
}%
\providecommand \@ifnum [1]{%
 \ifnum #1\expandafter \@firstoftwo
 \else \expandafter \@secondoftwo
 \fi
}%
\providecommand \@ifx [1]{%
 \ifx #1\expandafter \@firstoftwo
 \else \expandafter \@secondoftwo
 \fi
}%
\providecommand \natexlab [1]{#1}%
\providecommand \enquote  [1]{``#1''}%
\providecommand \bibnamefont  [1]{#1}%
\providecommand \bibfnamefont [1]{#1}%
\providecommand \citenamefont [1]{#1}%
\providecommand \href@noop [0]{\@secondoftwo}%
\providecommand \href [0]{\begingroup \@sanitize@url \@href}%
\providecommand \@href[1]{\@@startlink{#1}\@@href}%
\providecommand \@@href[1]{\endgroup#1\@@endlink}%
\providecommand \@sanitize@url [0]{\catcode `\\12\catcode `\$12\catcode
  `\&12\catcode `\#12\catcode `\^12\catcode `\_12\catcode `\%12\relax}%
\providecommand \@@startlink[1]{}%
\providecommand \@@endlink[0]{}%
\providecommand \url  [0]{\begingroup\@sanitize@url \@url }%
\providecommand \@url [1]{\endgroup\@href {#1}{\urlprefix }}%
\providecommand \urlprefix  [0]{URL }%
\providecommand \Eprint [0]{\href }%
\providecommand \doibase [0]{http://dx.doi.org/}%
\providecommand \selectlanguage [0]{\@gobble}%
\providecommand \bibinfo  [0]{\@secondoftwo}%
\providecommand \bibfield  [0]{\@secondoftwo}%
\providecommand \translation [1]{[#1]}%
\providecommand \BibitemOpen [0]{}%
\providecommand \bibitemStop [0]{}%
\providecommand \bibitemNoStop [0]{.\EOS\space}%
\providecommand \EOS [0]{\spacefactor3000\relax}%
\providecommand \BibitemShut  [1]{\csname bibitem#1\endcsname}%
\let\auto@bib@innerbib\@empty
\bibitem [{Note1()}]{Note1}%
  \BibitemOpen
  \bibinfo {note} {{\protect \color {black}Here we converted the summation in
  the first term of Eq.~\protect \textup {\hbox {\mathsurround \z@ \protect
  \normalfont (\ignorespaces \ref {gf}\unskip \@@italiccorr )}} into an
  integral~\protect \textup {\hbox {\mathsurround \z@ \protect \normalfont
  (\ignorespaces \ref {mu-lam}\unskip \@@italiccorr )}} and thus dropped the
  subscript $n$ from the matrix $C_n$ given in \protect \textup {\hbox
  {\mathsurround \z@ \protect \normalfont (\ignorespaces \ref {cn-eqn}\unskip
  \@@italiccorr )}}.}}\BibitemShut {Stop}%
\bibitem [{Note2()}]{Note2}%
  \BibitemOpen
  \bibinfo {note} {The subscript R indicates the right heat bath.}\BibitemShut
  {Stop}%
\end{thebibliography}%


\bigskip

\begin{thebibliography}{}
\bibitem{seifert-1} Seifert, U., 2012. Stochastic thermodynamics, fluctuation theorems and molecular machines. Reports on progress in physics, 75(12), p.126001.

\bibitem{van} Van Kampen, N.G., 1992. Stochastic processes in physics and chemistry (Vol. 1). Elsevier.

\bibitem{c-j-1} Klages, R., Just, W. and Jarzynski, C. eds., 2013. Nonequilibrium statistical physics of small systems. Wiley-VCH Verlag GmbH \& Company KGaA.

\bibitem{ret-1} Ritort, F., 2008. Nonequilibrium fluctuations in small systems: From physics to biology. Advances in chemical physics, 137, p.31.

\bibitem{eq-st} Plischke, M. and Bergersen, B., 2006. Equilibrium statistical physics. World Scientific Publishing Company.


\bibitem{ft-1} Evans, D.J., Cohen, E.G.D. and Morriss, G.P., 1993. Probability of second law violations in shearing steady states. Physical review letters, 71(15), p.2401.

\bibitem{ft-2} Searles, D.J. and Evans, D.J., 2000. Ensemble dependence of the transient fluctuation theorem. The Journal of Chemical Physics, 113(9), pp.3503-3509.

\bibitem{ft-3} Searles, D.J. and Evans, D.J., 2001. Fluctuation theorem for heat flow. International journal of thermophysics, 22(1), pp.123-134.

\bibitem{ft-4} Kurchan, J., 1998. Fluctuation theorem for stochastic dynamics. Journal of Physics A: Mathematical and General, 31(16), p.3719.

\bibitem{c-j-2} Jarzynski, C., 1997. Nonequilibrium equality for free energy differences. Physical Review Letters, 78(14), p.2690.

\bibitem{crooks} Crooks, G.E., 1999. Entropy production fluctuation theorem and the nonequilibrium work relation for free energy differences. Physical Review E, 60(3), p.2721.

\bibitem{HSR} Hatano, T. and Sasa, S.I., 2001. Steady-state thermodynamics of Langevin systems. Physical review letters, 86(16), p.3463.

\bibitem{tur-1} Barato, A.C. and Seifert, U., 2015. Thermodynamic uncertainty relation for biomolecular processes. Physical review letters, 114(15), p.158101.

\bibitem{Roldan-infer} {\color{black} Roldán, É., Neri, I., Dörpinghaus, M., Meyr, H. and Jülicher, F., 2015. Decision making in the arrow of time. Physical review letters, 115(25), p.250602.}

\bibitem{tur-2} Manikandan, S.K., Gupta, D. and Krishnamurthy, S., 2020. Inferring entropy production from short experiments. Physical review letters, 124(12), p.120603.

\bibitem{tur-3} Van Vu, T. and Hasegawa, Y., 2020. Entropy production estimation with optimal current. Physical Review E, 101(4), p.042138.

\bibitem{tur-4} Otsubo, S., Ito, S., Dechant, A. and Sagawa, T., 2020. Estimating entropy production by machine learning of short-time fluctuating currents. Physical Review E, 101(6), p.062106.

\bibitem{tur-5} Otsubo, S., Manikandan, S.K., Sagawa, T. and Krishnamurthy, S., 2020. Estimating entropy production along a single non-equilibrium trajectory. arXiv preprint arXiv:2010.03852.

\bibitem{ldf} Touchette, H., 2009. The large deviation approach to statistical mechanics. Physics Reports, 478(1-3), pp.1-69.

\bibitem{lde-1} Mehl, J., Speck, T. and Seifert, U., 2008. Large deviation function for entropy production in driven one-dimensional systems. Physical Review E, 78(1), p.011123.


\bibitem{lde-2} Gupta, D. and Sabhapandit, S., 2017. Stochastic efficiency of an isothermal work-to-work converter engine. Physical Review E, 96(4), p.042130.

\bibitem{lde-3} Verley, G., Esposito, M., Willaert, T. and Van den Broeck, C., 2014. The unlikely Carnot efficiency. Nature communications, 5(1), pp.1-5.

\bibitem{lde-4} Verley, G., Van den Broeck, C. and Esposito, M., 2014. Work statistics in stochastically driven systems. New Journal of Physics, 16(9), p.095001.

\bibitem{lde-5} Sabhapandit, S., 2011. Work fluctuations for a harmonic oscillator driven by an external random force. EPL (Europhysics Letters), 96(2), p.20005.

\bibitem{lde-6} Lacoste, D., Lau, A.W.C. and Mallick, K., 2008. Fluctuation theorem and large deviation function for a solvable model of a molecular motor. Physical Review E, 78(1), p.011915.

\bibitem{lde-7} Harris, R.J. and Touchette, H., 2009. Current fluctuations in stochastic systems with long-range memory. Journal of Physics A: Mathematical and Theoretical, 42(34), p.342001.

\bibitem{lde-8} Fischer, L.P., Pietzonka, P. and Seifert, U., 2018. Large deviation function for a driven underdamped particle in a periodic potential. Physical Review E, 97(2), p.022143.

\bibitem{am-1} Marchetti, M.C., Joanny, J.F., Ramaswamy, S., Liverpool, T.B., Prost, J., Rao, M. and Simha, R.A., 2013. Hydrodynamics of soft active matter. Reviews of Modern Physics, 85(3), p.1143.

\bibitem{am-2} Ramaswamy, S., 2010. The mechanics and statistics of active matter. Annu. Rev. Condens. Matter Phys., 1(1), pp.323-345.

\bibitem{am-3} Takatori, S.C. and Brady, J.F., 2015. Towards a thermodynamics of active matter. Physical Review E, 91(3), p.032117.

\bibitem{am-4} Ramaswamy, S., 2017. Active matter. Journal of Statistical Mechanics: Theory and Experiment, 2017(5), p.054002.

\bibitem{am-5} J$\ddot{\text{u}}$licher, F., Grill, S.W. and Salbreux, G., 2018. Hydrodynamic theory of active matter. Reports on Progress in Physics, 81(7), p.076601.

\bibitem{am-6} De Magistris, G. and Marenduzzo, D., 2015. An introduction to the physics of active matter. Physica A: Statistical Mechanics and its Applications, 418, pp.65-77.

\bibitem{am-7} Fodor, $\acute{\rm E}$ . and Marchetti, M.C., 2018. The statistical physics of active matter: From self-catalytic colloids to living cells. Physica A: Statistical Mechanics and its Applications, 504, pp.106-120.

\bibitem{am-8} Schweitzer, F., 2003. Brownian agents and active particles: collective dynamics in the natural and social sciences. Springer Science \& Business Media.



\bibitem{fsc-1} Vicsek, T., Czirók, A., Ben-Jacob, E., Cohen, I. and Shochet, O., 1995. Novel type of phase transition in a system of self-driven particles. Physical review letters, 75(6), p.1226.

\bibitem{am-11} Toner, J. and Tu, Y., 1998. Flocks, herds, and schools: A quantitative theory of flocking. Physical review E, 58(4), p.4828.

\bibitem{mod} Romanczuk, P., Bär, M., Ebeling, W., Lindner, B. and Schimansky-Geier, L., 2012. Active brownian particles. The European Physical Journal Special Topics, 202(1), pp.1-162.

\bibitem{fsc-2} Hubbard, S., Babak, P., Sigurdsson, S.T. and Magnússon, K.G., 2004. A model of the formation of fish schools and migrations of fish. Ecological Modelling, 174(4), pp.359-374.

\bibitem{fl-0} Cavagna, A. and Giardina, I., 2014. Bird flocks as condensed matter. Annu. Rev. Condens. Matter Phys., 5(1), pp.183-207.

\bibitem{fl-1} Toner, J., Tu, Y. and Ramaswamy, S., 2005. Hydrodynamics and phases of flocks. Annals of Physics, 318(1), pp.170-244.

\bibitem{fl-2} Kumar, N., Soni, H., Ramaswamy, S. and Sood, A.K., 2014. Flocking at a distance in active granular matter. Nature communications, 5(1), pp.1-9. 

\bibitem{csf} Palacci, J., Sacanna, S., Steinberg, A.P., Pine, D.J. and Chaikin, P.M., 2013. Living crystals of light-activated colloidal surfers. Science, 339(6122), pp.936-940.

\bibitem{ecoli} Berg, H.C., 2008. E. coli in Motion. Springer Science \& Business Media.

\bibitem{bac} Cates, M.E., 2012. Diffusive transport without detailed balance in motile bacteria: does microbiology need statistical physics?. Reports on Progress in Physics, 75(4), p.042601.

\bibitem{syn-1} Ozin, G.A., Manners, I., Fournier-Bidoz, S. and Arsenault, A., 2005. Dream nanomachines. Advanced Materials, 17(24), pp.3011-3018.

\bibitem{syn-2} Dreyfus, R., Baudry, J., Roper, M.L., Fermigier, M., Stone, H.A. and Bibette, J., 2005. Microscopic artificial swimmers. Nature, 437(7060), pp.862-865.

\bibitem{syn-3} Volpe, G., Buttinoni, I., Vogt, D., Kümmerer, H.J. and Bechinger, C., 2011. Microswimmers in patterned environments. Soft Matter, 7(19), pp.8810-8815.

\bibitem{mot-c} Comelles, J., Caballero, D., Voituriez, R., Hortigüela, V., Wollrab, V., Godeau, A.L., Samitier, J., Martínez, E. and Riveline, D., 2014. Cells as active particles in asymmetric potentials: motility under external gradients. Biophysical journal, 107(7), pp.1513-1522.

\bibitem{cl-1} Redner, G.S., Hagan, M.F. and Baskaran, A., 2013. Structure and dynamics of a phase-separating active colloidal fluid. Physical review letters, 110(5), p.055701.

\bibitem{cl-2} Bricard, A., Caussin, J.B., Desreumaux, N., Dauchot, O. and Bartolo, D., 2013. Emergence of macroscopic directed motion in populations of motile colloids. Nature, 503(7474), pp.95-98.

\bibitem{press} Solon, A.P., Fily, Y., Baskaran, A., Cates, M.E., Kafri, Y., Kardar, M. and Tailleur, J., 2015. Pressure is not a state function for generic active fluids. Nature Physics, 11(8), pp.673-678.

\bibitem{mips} Cates, M.E. and Tailleur, J., 2015. Motility-induced phase separation. Annu. Rev. Condens. Matter Phys., 6(1), pp.219-244.

\bibitem{jam} Slowman, A.B., Evans, M.R. and Blythe, R.A., 2016. Jamming and attraction of interacting run-and-tumble random walkers. Physical review letters, 116(21), p.218101.


\bibitem{sfd-1} Teomy, E. and Metzler, R., 2019. Transport in exclusion processes with one-step memory: density dependence and optimal acceleration. Journal of Physics A: Mathematical and Theoretical, 52(38), p.385001.

\bibitem{sfd-2} Teomy, E. and Metzler, R., 2019. Correlations and transport in exclusion processes with general finite memory. Journal of Statistical Mechanics: Theory and Experiment, 2019(10), p.103211.

\bibitem{sfd-3} Galanti, M., Fanelli, D. and Piazza, F., 2013. Persistent random walk with exclusion. The European Physical Journal B, 86(11), pp.1-5.

\bibitem{sfd-4} Dolai, P., Das, A., Kundu, A., Dasgupta, C., Dhar, A. and Kumar, K.V., 2020. Universal scaling in active single-file dynamics. Soft Matter, 16(30), pp.7077-7087.

\bibitem{res-1} Evans, M.R. and Majumdar, S.N., 2018. Run and tumble particle under resetting: a renewal approach. Journal of Physics A: Mathematical and Theoretical, 51(47), p.475003.

\bibitem{res-2} Kumar, V., Sadekar, O. and Basu, U., 2020. Active Brownian motion in two dimensions under stochastic resetting. Physical Review E, 102(5), p.052129.

\bibitem{pd-1} Basu, U., Majumdar, S.N., Rosso, A. and Schehr, G., 2019. Long-time position distribution of an active Brownian particle in two dimensions. Physical Review E, 100(6), p.062116.

\bibitem{pd-2} Basu, U., Majumdar, S.N., Rosso, A., Sabhapandit, S. and Schehr, G., 2020. Exact stationary state of a run-and-tumble particle with three internal states in a harmonic trap. Journal of Physics A: Mathematical and Theoretical, 53(9), p.09LT01.

\bibitem{pd-3} Das, S., Gompper, G. and Winkler, R.G., 2018. Confined active Brownian particles: theoretical description of propulsion-induced accumulation. New Journal of Physics, 20(1), p.015001.

\bibitem{surp} Mori, F., Le Doussal, P., Majumdar, S.N. and Schehr, G., 2020. Universal survival probability for a d-dimensional run-and-tumble particle. Physical review letters, 124(9), p.090603.

\bibitem{ak} Singh, P. and Kundu, A., 2020. Correlation and fluctuation in chain of active particles. arXiv preprint arXiv:2012.13910.

\bibitem{sptm}Caprini, L. and Marconi, U.M.B., 2020. Time-dependent properties of interacting active matter: Dynamical behavior of one-dimensional systems of self-propelled particles. Physical Review Research, 2(3), p.033518.

\bibitem{arc} Singh, P. and Kundu, A., 2019. Generalised 'Arcsine' laws for run-and-tumble particle in one dimension. Journal of Statistical Mechanics: Theory and Experiment, 2019(8), p.083205.


\bibitem{conh}Hartmann, A.K., Majumdar, S.N., Schawe, H. and Schehr, G., 2020. The convex hull of the run-and-tumble particle in a plane. Journal of Statistical Mechanics: Theory and Experiment, 2020(5), p.053401.


\bibitem{abp-rtp} Solon, A.P., Cates, M.E. and Tailleur, J., 2015. Active brownian particles and run-and-tumble particles: A comparative study. The European Physical Journal Special Topics, 224(7), pp.1231-1262.

\bibitem{aoup} Martin, D., O'Byrne, J., Cates, M.E., Fodor, É., Nardini, C., Tailleur, J. and van Wijland, F., 2020. Statistical mechanics of active ornstein uhlenbeck particles. arXiv preprint arXiv:2008.12972.

\bibitem{ion} Santra, I., Basu, U. and Sabhapandit, S., 2021. Active Brownian Motion with Directional Reversals. arXiv preprint arXiv:2101.11327.

\bibitem{glass} Szamel, G., Flenner, E. and Berthier, L., 2015. Glassy dynamics of athermal self-propelled particles: Computer simulations and a nonequilibrium microscopic theory. Physical Review E, 91(6), p.062304.


\bibitem{walls} Caprini, L. and Marconi, U.M.B., 2018. Active particles under confinement and effective force generation among surfaces. Soft matter, 14(44), pp.9044-9054.


\bibitem{ldsbo-1} Dabelow, L., Bo, S. and Eichhorn, R., 2020. How irreversible are steady-state trajectories of a trapped active particle?. arXiv preprint arXiv:2012.05542.

\bibitem{ldsbo-2} Fodor, \'E., Nardini, C., Cates, M.E., Tailleur, J., Visco, P. and van Wijland, F., 2016. How far from equilibrium is active matter?. Physical review letters, 117(3), p.038103.

\bibitem{ldsbo-3} Caprini, L., Marconi, U.M.B., Puglisi, A. and Vulpiani, A., 2019. The entropy production of Ornstein–Uhlenbeck active particles: a path integral method for correlations. Journal of Statistical Mechanics: Theory and Experiment, 2019(5), p.053203.

\bibitem{ht-1} Kundu, A., Sabhapandit, S. and Dhar, A., 2011. Large deviations of heat flow in harmonic chains. Journal of Statistical Mechanics: Theory and Experiment, 2011(03), p.P03007.

\bibitem{Fogedby}Fogedby, H.C. and Imparato, A., 2012. Heat flow in chains driven by thermal noise. Journal of Statistical Mechanics: Theory and Experiment, 2012(04), p.P04005.


\bibitem{ht-2} Dhar, A. and Dandekar, R., 2015. Heat transport and current fluctuations in harmonic crystals. Physica A: Statistical Mechanics and its Applications, 418, pp.49-64.

\bibitem{ht-3} Dhar, A. and Roy, D., 2006. Heat transport in harmonic lattices. Journal of Statistical Physics, 125(4), pp.801-820.

\bibitem{ht-4} Dhar, A. and Saito, K., 2016. Heat transport in harmonic systems. In Thermal Transport in Low Dimensions (pp. 39-105). Springer, Cham.

\bibitem{one-two-har-anhar} Lepri, S., Livi, R. and Politi, A., 2003. Thermal conduction in classical low-dimensional lattices. Physics reports, 377(1), pp.1-80.

\bibitem{non-li} Lepri, S., Livi, R. and Politi, A., 1997. Heat conduction in chains of nonlinear oscillators. Physical review letters, 78(10), p.1896.
\bibitem{ht-6} Das, S.G., Dhar, A. and Narayan, O., 2014. Heat conduction in the $\alpha-\beta$ Fermi–Pasta–Ulam chain. Journal of Statistical Physics, 154(1), pp.204-213.



\bibitem{ht-5} Saito, K. and Dhar, A., 2010. Heat conduction in a three dimensional anharmonic crystal. Physical review letters, 104(4), p.040601.





\bibitem{ht-7} Dhar, A. and Saito, K., 2008. Heat conduction in the disordered Fermi-Pasta-Ulam chain. Physical Review E, 78(6), p.061136.

\bibitem{Fogedby-0}Fogedby, H.C., 2014. Large deviations in the alternating mass harmonic chain. Journal of Physics A: Mathematical and Theoretical, 47(32), p.325003.


\bibitem{ht-8} Dhar, A., 2001. Heat conduction in a one-dimensional gas of elastically colliding particles of unequal masses. Physical review letters, 86(16), p.3554.

\bibitem{ht-9} Sabhapandit, S., 2012. Heat and work fluctuations for a harmonic oscillator. Physical Review E, 85(2), p.021108.


\bibitem{Fogedby-2}Fogedby, H.C. and Imparato, A., 2011. A bound particle coupled to two thermostats. Journal of Statistical Mechanics: Theory and Experiment, 2011(05), p.P05015.

\bibitem{Visco}Visco, P., 2006. Work fluctuations for a Brownian particle between two thermostats. Journal of Statistical Mechanics: Theory and Experiment, 2006(06), p.P06006.

\bibitem{apal} Pal, A. and Sabhapandit, S., 2014. Work fluctuations for a Brownian particle driven by a correlated external random force. Physical Review E, 90(5), p.052116.

\bibitem{sekimoto} Sekimoto, K., 1998. Langevin equation and thermodynamics. Progress of Theoretical Physics Supplement, 130, pp.17-27.

\bibitem{ito} Van Kampen, N.G., 1981. It\^o versus stratonovich. Journal of Statistical Physics, 24(1), pp.175-187.


\bibitem{apal-2} Pal, A. and Sabhapandit, S., 2013. Work fluctuations for a Brownian particle in a harmonic trap with fluctuating locations. Physical Review E, 87(2), p.022138.

\bibitem{pep-1} Gupta, D. and Sabhapandit, S., 2016. Fluctuation theorem for entropy production of a partial system in the weak-coupling limit. EPL (Europhysics Letters), 115(6), p.60003.

\bibitem{pep-2} Gupta, D. and Sabhapandit, S., 2018. Partial entropy production in heat transport. Journal of Statistical Mechanics: Theory and Experiment, 2018(6), p.063203.

\bibitem{pep-3} Gupta, D. and Sabhapandit, S., 2020. Entropy production for partially observed harmonic systems. Journal of Statistical Mechanics: Theory and Experiment, 2020(1), p.013204.


\bibitem{Lebowitz}Bonetto, F., Lebowitz, J.L. and Lukkarinen, J., 2004. Fourier's law for a harmonic crystal with self-consistent stochastic reservoirs. Journal of statistical physics, 116(1-4), pp.783-813.

\bibitem{Falasco}Falasco, G., Baiesi, M., Molinaro, L., Conti, L. and Baldovin, F., 2015. Energy repartition for a harmonic chain with local reservoirs. Physical Review E, 92(2), p.022129.

\bibitem{adhar-review} {\color{black}Dhar, A., 2008. Heat transport in low-dimensional systems. Advances in Physics, 57(5), pp.457-537.}


\bibitem{mmeth} Arfken, G.B. and Weber, H.J., 1999. Mathematical methods for physicists.

\bibitem{Kardar}Kardar, M., 2007. Statistical physics of particles. Cambridge University Press.


\end{thebibliography}
\end{document}